# Benchmarking foundation models as feature extractors for weakly-supervised computational pathology

Peter Neidlinger (1,*), Omar S. M. El Nahhas (1, 2, *), Hannah Sophie Muti (1, 3, 4), Tim Lenz (1), Michael Hoffmeister (5), Hermann Brenner (5, 6, 7), Marko van Treeck (1), Rupert Langer (8), Bastian Dislich (9), Hans Michael Behrens (10), Christoph Röcken (10), Sebastian Foersch (11), Daniel Truhn (2, 12), Antonio Marra (13), Oliver Lester Saldanha (1), Jakob Nikolas Kather (1, 2, 4, 14, 15, +)

* equal contribution
\+ Correspondence to jakob-nikolas.kather@alumni.dkfz.de

1. Else Kroener Fresenius Center for Digital Health, Technical University Dresden, Dresden, Germany
2. StratifAI GmbH, Dresden, Germany
3. Department for Visceral, Thoracic and Vascular Surgery, University Hospital and Faculty of Medicine Carl Gustav Carus, Technische Universität Dresden, Dresden, Germany
4. National Center for Tumor Diseases Dresden (NCT/UCC), a partnership between DKFZ, Faculty of Medicine and University Hospital Carl Gustav Carus, TUD Dresden University of Technology, and Helmholtz-Zentrum Dresden - Rossendorf (HZDR), Dresden, Germany
5. Division of Clinical Epidemiology and Aging Research, German Cancer Research Center (DKFZ), Heidelberg, Germany
6. Division of Preventive Oncology, German Cancer Research Center (DKFZ) and National Center for Tumor Diseases (NCT), Heidelberg, Germany
7. German Cancer Consortium (DKTK), German Cancer Research Center (DKFZ), Heidelberg, Germany
8. Institute of Pathology and Molecular Pathology, Kepler University Hospital, Johannes Kepler University Linz, Linz, Austria
9. Institute of Tissue Medicine and Pathology, University of Bern, Bern, Switzerland
10. Department of Pathology, University Hospital Schleswig-Holstein, Kiel, Germany
11. Institute of Pathology, University Medical Center Mainz, Mainz, Germany
12. Department of Diagnostic and Interventional Radiology, University Hospital Aachen
13. Division of New Drugs and Early Drug Development, European Institute of Oncology IRCCS, Milan, Italy
14. Medical Department 1, University Hospital and Faculty of Medicine Carl Gustav Carus, Technische Universität Dresden, Dresden, Germany
15. Medical Oncology, National Center for Tumor Diseases (NCT), University Hospital Heidelberg, Heidelberg, Germany

# Abstract

Advancements in artificial intelligence have driven the development of numerous pathology foundation models capable of extracting clinically relevant information. However, there is currently limited literature independently evaluating these foundation models on truly external cohorts and clinically-relevant tasks to uncover adjustments for future improvements. In this study, we benchmarked 19 histopathology foundation models on 13 patient cohorts with 6,818 patients and 9,528 slides from lung, colorectal, gastric, and breast cancers. The models were evaluated on weakly-supervised tasks related to biomarkers, morphological properties, and prognostic outcomes. We show that a vision-language foundation model, CONCH, yielded the highest performance when compared to vision-only foundation models, with Virchow2 as close second. The experiments reveal that foundation models trained on distinct cohorts learn complementary features to predict the same label, and can be fused to outperform the current state of the art. An ensemble combining CONCH and Virchow2 predictions outperformed individual models in 55% of tasks, leveraging their complementary strengths in classification scenarios. Moreover, our findings suggest that data diversity outweighs data volume for foundation models. Our work highlights actionable adjustments to improve pathology foundation models.

# Introduction

Artificial intelligence (AI) has revolutionized digital pathology (DP) by enabling biomarker prediction from cancer tissues using high-resolution whole slide images (WSIs) [1–6]. Moreover, these algorithms can substantially enhance diagnostic accuracy, efficiency and consistency, significantly reducing the subjectivity associated with human interpretation [7,8]. In particular, deep learning (DL) can perform tasks such as disease grading, cancer subclassification or prognostic prediction [9–11].

Recently, foundation models, which are trained on large-scale datasets, have been introduced to DP [12,13]. These models use self-supervised Learning (SSL) techniques to learn meaningful representations of histology tissue, which are crucial for clinical pathology tasks. SSL techniques such as contrastive learning [14,15] and masked image modeling (MIM) [16] have shown improved performance, robustness and higher transferability compared to fully supervised learning. Another advantage lies in its ability to learn from vast amounts of unlabeled data, thereby significantly reducing the need for manual annotation [17]. The practical application of foundation models involves WSI tessellation into small, non-overlapping patches, after which image feature extraction is performed. These extracted features serve as inputs for training classification or regression models, such as ViTs [18], tailored for specific tasks, like mutation prediction, survival analysis, disease grading, or cancer classification [19]. The limited availability and variable quality of public pathology data can hinder the performance of these models when applied to real-world clinical scenarios [20]. Recent efforts have demonstrated the potential of large-scale foundation models in computational pathology. Unlike earlier models that relied heavily on datasets like The Cancer Genome Atlas (TCGA), contemporary foundation models are now trained on much larger proprietary cohorts like Mass-100K (100k WSIs) [21], Providence (171k WSIs) [22] and Memorial Sloan Kettering Cancer Center (1488k WSIs) [23].

Foundation models have enabled the rapid development of specialized, task-specific downstream models by providing a stable base architecture. These downstream models require substantially less data and computational resources since they build upon the pre-existing foundation model. While the success of foundation models is typically measured by downstream model performance, their evaluation has largely been limited to narrow benchmarks without proper external validation. This restricted testing approach risks data leakage and selective reporting of only the best-performing models. As a result, most foundation models lack systematic evaluation across a broad spectrum of clinically relevant tasks, leaving their true capabilities and limitations incompletely understood.

In this study, we put forth a comprehensive benchmarking effort for histopathology foundation models. By including multiple proprietary cohorts from multiple countries, which were never part of any foundation model training, we effectively mitigate the risk of data leakage from pretraining datasets. Our benchmarking includes 19 foundation models and 31 clinically relevant evaluation tasks, 19 of which are the prediction of cancer biomarkers, using a total of 6,818 patients and 9,528 slides. This comprehensive evaluation bridges a notable gap in DP literature and will serve as an important reference point for the DP community helping to select the right foundation model for a specific DP task.

# Results

## Benchmark of pathology foundation models

We benchmarked the performance of 19 foundation models and 14 ensembles derived from these models, trained as vision-language or vision-only, on 31 weakly-supervised downstream prediction tasks related to morphology (n=5), biomarkers (n=19), and prognostication (n=7).

For the five morphology-related tasks, CONCH yielded the highest mean AUROC of 0.77, followed by Virchow2 and DinoSSLPath with mean AUROCs of 0.76 (**Figure 2C**). Across the 19 biomarker-related tasks, Virchow2 and CONCH achieved the highest mean AUROCs of 0.73, followed closely by Prov-GigaPath with a mean AUROC of 0.72 (**Figure 2D**). Finally, in the seven prognostic-related tasks, CONCH yielded the highest mean AUROC of 0.63, followed by Virchow2 and BiomedCLIP with mean AUROCs of 0.61 (**Figure 2E**). Averaged across all 31 tasks, CONCH and Virchow2 had the highest AUROCs of 0.71, followed by Prov-GigaPath and DinoSSLPath with AUROCs of 0.69. Subsequent rankings included H-optimus-0, UNI and Panakeia (0.68), Virchow, Hibou-L and CTransPath (0.67), BiomedCLIP and Kaiko (0.66), Phikon (0.65) and PLIP (0.64). Moreover, CONCH achieved the highest average area under the precision recall characteristic (AUPRC), balanced accuracy and F1 scores (**Figure S1**), with the highest average AUROC in each cancer type obtained by CONCH (STAD, NSCLC), Virchow2 (CRC) and BiomedCLIP (BRCA). (**Figure S2A**). To further validate our findings, we compared the performance of transformer-based aggregation with the widely used Attention-Base Multiple Instance Learning (ABMIL) approach [24]. Across all 31 tasks, ABMIL performed slightly worse than the transformer-based model, with an average AUROC difference of 0.01, leaving the overall model rankings largely unchanged (**Figure S3**).

For histopathology slide encoders, we retrieved the encoded tile-level embeddings to make them applicable to our MIL approach. The original tile embeddings consistently outperformed their slide-level counterparts and the performance of the encoded tile embeddings is driven by the quality of the original tile embeddings and not by the slide encoder (**Figure 2G**).

In statistical AUROC comparisons across 29 binary classification tasks, CONCH yielded higher AUROCs which were significantly different from other models in a substantial number of tasks: PLIP (16), Phikon and BiomedCLIP (13), Kaiko (11), and 7 tasks each for Hibou-L, H-optimus-0, CTransPath, Virchow, Panakeia, UNI, and DinoSSLPath, with 5 tasks each for Prov-GigaPath and Virchow2. Conversely, few models yielded higher AUROCs than CONCH: Virchow2 (6), Prov-GigaPath (3), Panakeia and Kaiko (2), and DinoSSLPath, UNI, Virchow and Hibou-L (1). Notably, PLIP, Phikon, BiomedCLIP, H-optimus-0 and CTransPath were not significantly better than CONCH in any of the tasks ($p < 0.05$; **Figure S4B**). Among the vision-only models, Virchow2 was significantly better than all other models in between 6 and 12 tasks ($p < 0.05$; **Figure S4C**).

Together, these data show that CONCH, a vision-language model trained on 1.17 million image-caption pairs (ICPs), performs on par with Virchow2, a vision-only model trained on 3.1 million WSIs, and together outperform all other pathology foundation models in the three highlighted domains of morphology, biomarkers and prognostication-based prediction tasks and that slide encoders are ineffective in a MIL setup.

## Performance of pathology foundation models in scarce data settings

One of the predominant selling points of foundation models in computational pathology is the mitigation of the traditional requirement for extensive labeled datasets when analyzing rare (molecular) events. Consequently, we analyzed the performance of pathology foundation models across two dimensions: WSI count for foundation model training, and patient and positive case counts for downstream model training, with emphasis on low-prevalence scenarios that reflect real-world clinical applications.

From the foundation model perspective, positive correlations ($r=0.29$ to $0.74$) were observed between downstream performance and pretraining dataset size (WSIs, patients) or diversity (tissue sites) across morphology, biomarker, and prognosis tasks, though most were not statistically significant. Significant correlations were found only for morphology with patient count ($r=0.73$, $p<0.05$) and tissue site diversity ($r=0.74$, $p<0.05$) (**Figure 3A**). These findings suggest these factors are important but not sole determinants, with the distribution of anatomic tissue sites (**Table S1, Figure S5**), architecture and dataset quality also playing critical roles. This is especially evident in vision-language models, where CONCH outperformed BiomedCLIP despite seeing far fewer ICPs (1.1M vs. 15M) (**Figure 3B**). Similarly, tissue representation in pretraining datasets showed a moderate, but not significant, correlation with performance by cancer type (**Figure 3C**). Interestingly, Panakeia models showed decent performance on unrelated cancer types, with the BRCA model achieving average results in NSCLC and the CRC model performing similarly in STAD, despite no prior exposure to these tissues during training.

Downstream models were trained on randomly sampled cohorts of 300, 150, and 75 patients while keeping a similar ratio of positive samples, and consequently validated on full-size external cohorts. In the largest sampled cohort (n=300), Virchow2 demonstrated superior performance in 8 tasks, followed closely by PRISM with 7 tasks. With the medium-sized sampled cohort (n=150), PRISM dominated by leading in 9 tasks, while Virchow2 followed with 6 tasks. The smallest sampled cohort size (n=75) showed more balanced results, with CONCH leading in 5 tasks, while PRISM and Virchow2 each led in 4 tasks. Performance metrics remained relatively stable between n=75 and n=150 cohorts (**Figure 3D, 3E, S6**).

To evaluate foundation models in real-world clinical scenarios, we focused on clinically relevant tasks with rare positive cases (>15%) in the TCGA training cohort. Key low-prevalence biomarkers included BRAF mutation (10%), CIMP status (13%), and MSI status (14%) in CRC; EBV positivity (8%) and M-status (7%) in STAD; and EGFR mutation (11%) and STK11 mutation (15%) in LUAD. To avoid cancer type imbalance, these targets were only evaluated in DACHS, Kiel and CPTAC-LUAD. The results show that Prov-GigaPath (mean AUROC of 0.74) yields the highest performance in the highlighted low-prevalence tasks, followed by Virchow (0.73) and CONCH (0.72) (**Figure S2B**).

Finally, tasks were stratified into high- and low-performance tasks by the AUROC (**Figure S7**). In high-performance tasks (>0.75), Virchow2 demonstrated superior performance in high-performance tasks, followed by Prov-GigaPath and CONCH. Conversely, in low-performance tasks (≤0.75), CONCH yielded better results.

Together, these results indicate that the patient count, tissue site diversity, and their distribution are important for downstream performance, though other factors like architecture and dataset quality also play critical roles. Moreover, the performance in downstream tasks with low-prevalence cases indicates the limitations of current foundation models for nonetheless clinically-relevant biomarkers . Lastly, we show differential model efficacy based on task complexity, with Virchow2 excelling in standard classification tasks while CONCH predominates in more challenging predictive scenarios. All models show similar performance declines with reduced training sizes, underlining the weakness of current pathology foundation models in scarce data scenarios.

## Pathology foundation models learn different tissue morphologies

To quantitatively measure prediction similarity across models, we calculated Cohen's kappa[25]. For each task, labels were assigned using a majority vote across the cross-validation folds. Cohen’s kappa scores were generally moderate and varied across models. Notably, some pairs such as Panakeia and UNI (0.66), PLIP and BiomedCLIP (0.52) and top performers like Prov-GigaPath, CONCH, Virchow2, and DinoSSLPath showed higher agreement, whereas lower-performing models such as Hibou and Kaiko exhibited the least consensus (0.28) (**Figure 4B**). Within individual model folds, BiomedCLIP and CONCH achieved the highest average kappa (0.41), followed by Virchow2, Panakeia and Prov-GigaPath (0.37), with Hibou (0.26) and Kaiko (0.24) ranking lowest, consistent with their AUROC performance (**Figure 4C**).

To elucidate the reasons behind the observed performance differences among the downstream models trained on top of the different foundation models, we investigated whether the models focus on different morphological properties for their predictions. We utilized attention heatmaps to compare model behavior when the models 1) consistently predicted the label correctly and 2) were in disagreement regarding the predicted label. In cases where all models were in agreement on the correct prediction, the validity of the classification would be supported by their focus on relevant tissue regions for diagnosis. For example, in the prediction of MSI status, models predominantly highlighted tumor regions, as expected. However, models such as UNI, Hibou, Virchow and Kaiko occasionally highlighted pen marks, which is an undesired behavior that suggests predictions are being made through some form of pattern association rather than understanding the underlying biology (**Figure 4A, S8B**). To assess the impact of pen marks, we quantified their occurrence in 50 randomly sampled slides per test cohort and found them present in 90% of slides from DACHS and 22% from Bern, but absent elsewhere. Despite their presence, pen marks did not skew classification, as they were equally distributed across different classes. Models such as CONCH and Virchow focused on multiple small tissue areas, whereas Prov-GigaPath appears less selective in its attention (**Figure 4A**). In NSCLC subtyping, models generally performed well, focusing mainly on tumor regions and ignoring healthy lung parenchyma (**Figure S9B**). In *ESR1* overexpression prediction, Prov-GigaPath and Kaiko highlighted the majority of the WSI area, whereas CONCH and Virchow focused on a few small tissue areas (**Figure S9C**). In contrast, when analyzing slides where models made inconsistent predictions, we found instances of model disagreement that led to errors. For instance, in the task of DACHS CRC sidedness, Virchow erroneously focused on pen marks (**Figure S8B**). However, no consistent pattern of errors emerged across the models to fully explain these discrepancies.

Together, these data indicate that foundation models vary in their focus on tissue regions and the morphological features they prioritize, which impacts their predictive performance. The differences in attention across models suggest that combining models with complementary strengths could enhance overall predictive accuracy in ensemble approaches.

## Ensemble of pathology foundation models improve performance

Lastly, we tested the hypothesis that creating an ensemble of pathology foundation models improves prediction performance. We utilized two approaches for ensembling models, taking the average of the various downstream models' prediction scores trained on different foundation model backbones, and concatenating feature vectors from different foundation model backbones to create a single downstream model.

Experiments show that ensembling by taking the average of the models' prediction scores yielded a superior AUROC compared to either model used in isolation. The combination of the four top-performing models led to the highest improvement, achieving a mean AUROC 1.2% higher than CONCH (**Figure S10**), the leading individual model (**Figure 1B**). Across all 31 tasks, the Ensemble reduced misclassifications compared to CONCH by an average of 6.2% across the five folds (cut-off 0.5) (**Table S2**). Therefore, these data show that ensembling the prediction scores of multiple high-performing models enhances performance on certain tasks beyond the capabilities of the best individual model.

Combining the best-performing models, CONCH and Virchow2, yielded a 1792-dimensional vector with the highest AUROC of 71.9. Similarly, combining Virchow2 and Prov-GigaPath, the top-performing vision-only models, resulted in a 2816-dimensional vector with an AUROC of 71.6. Individually, the models achieved AUROCs of 71.1 for CONCH, 70.9 for Virchow2, and 69.2 for Prov-GigaPath (**Figure 1B, S10**). Interestingly, Cohen's kappa between the individual models did not strongly correlate with ensemble quality, indicating that low agreement does not necessarily translate to beneficial diversity in predictions. Similarly, no clear pattern was observed between the similarity of ensembles with their single model counterparts and factors like model performance or embedding size (**Figure S11**). To quantify improvements, we conducted two-sided DeLong's tests comparing AUROC scores of CONCH with ensembles and other single-model baselines. For each model, we averaged prediction scores across five folds, and across up to 10 folds for ensembles. Bagging the five folds of the same foundation model increased AUROC scores, while integrating different models via stacking or concatenation yielded more pronounced improvements (**Figure S4A**). The CONCH and Virchow2 ensemble showed statistically significant differences in performance with higher AUROCs than CONCH in nine of 29 tasks ($p<0.05$), whereas the Virchow2 and Prov-GigaPath ensemble showed significant improvements in seven tasks (**Figure S4B**).

These results demonstrate that ensemble approaches for pathology foundation models, as well as their downstream models, lead to enhanced prediction performance. This suggests that merging multiple foundation models through ensemble techniques can be beneficial.

# Discussion

Weakly-supervised computational pathology approaches, in which a deep learning system predicts a label directly from a whole slide image, have been massively successful in cancer research. They have been used to make the diagnosis of tumors, to predict biomarker status, and to predict clinical outcomes directly from image data. Over one hundred such tools are now approved for clinical use in the US and the European Union [26,27]. Since 2022, foundation models have become an integral part of weakly supervised computational pathology pipelines and have improved performance and generalizability [4,28]. However, the current internal evaluation strategy for foundation models in computational pathology for clinically relevant tasks is limited. When groups that publish pathology foundation models evaluate them on tasks of their own choosing, there is a high potential for bias. Moreover, concerns about data leakage arise when foundation models are tested on images from the same institutions where they were trained.

In this study, we conducted a comprehensive evaluation of pathology foundation models in weakly-supervised computational pathology on truly external datasets with no overlap between training and validation data. Our results show that while many existing foundation models achieve high performance on clinically relevant prediction tasks, CLIP-based approaches aren't inherently superior, as evidenced by BiomedCLIP and PLIP's performance. Rather, high-quality pre-training data and effective data cleaning are crucial for achieving top-tier performance. The best performing model, CONCH, trained with multimodal data, suggests that incorporating text during training enhances image-only embedding quality. Similarly, Virchow2's strong performance stems from its unprecedented tissue type diversity (approximately 200, versus 20-30 in other models) and more balanced distribution, avoiding over-representation of specific cancer types. Additionally, the variability in the model's performance can also be attributed to varying degrees of difficulty for each task. For instance, while differentiating between lung carcinoma subtypes is generally straightforward, other tasks like stomach cancer subtyping can be more demanding. Here, even pathologists can show a considerate degree of interobserver disagreement[29].

In terms of prediction interpretability, our approach highlights that different foundation models focus on different areas in the tissue, while still having a high agreement on the predicted label. Our technical analysis revealed that slide encoders showed no advantage over tile encoders in MIL setups, except in low-data scenarios, and the transformer-based STAMP architecture generally outperformed ABMIL outside of data-limited settings. We demonstrate that ensembling foundation models is beneficial, particularly when combining top-performing models, though prediction diversity (measured by Cohen's Kappa) doesn't directly correlate with ensemble performance. Even modest ensemble improvements may have clinical relevance by combining several learned perspectives of tissue morphology, as exemplified by the higher biomarker classification performance. Future work should incorporate more sophisticated methods than feature vector concatenation, especially for larger models where combining large vectors might lead to overfitting.

A key insight of our study is that performance of foundation models does not scale well with increasing numbers of images in the training set used for self-supervised learning. Meaning, bigger is not always better. Rather, the diversity of the training set suggests to be a key factor, favoring various sources of data, races, and types of cancer. Our results will inform the future development of new foundation models. Specifically, using multimodal data to train models, even if the intention is just to apply them on unimodal data (i.e. on images alone), should be encouraged. For healthcare institutions, this means that data which is available at scale, even without clinical association with clinical endpoints, is a valuable resource to train such models.

Our study has limitations in that our evaluation tasks only contain certain tumor types. We focused on four cancer types, prioritizing truly external validation datasets over broader cancer type coverage. This differentiates our work from studies that train and test on the same cohort or WSIs from the same hospital used for pretraining. Moreover, we were limited to pathology foundation models licenses which are accessible in a research setting. For example, this excludes RudolphV and PLUTO from our analysis. While our datasets contained artifacts like pen marks (present in 90% of DACHS and 22% of Bern samples), these had minimal impact on predictions due to their even distribution across classes. Though we incorporated a broad range of foundation models applicable to histology data, exploring the potential of fine-tuning general-purpose models like GPT-4o was outside our current scope. Our evaluation strategy is focused on a diverse set of biomarkers in cancer histopathology. Future work will expand upon the range of tumor types, biomarkers and patient cohorts to further evaluate the robustness of foundation models in pathology.

# Author contributions

PN, OSMEN, HSM and JNK designed the study. PN, TL, OSMEN and MVT developed the software. PN, MH, HB, HSM, RL, BD, HMB, CR, AM, OLS and JNK contributed to data collection and assembly. PN, OSMEN, TL, HSM, SF, DT and OLS interpreted and analyzed the data. All authors substantially contributed to writing and reviewing the report, approved the final version for submission, and have agreed to be personally accountable for the author's own contributions and to ensure that questions related to the accuracy or integrity of any part of the work, even ones in which the author was not personally involved, are appropriately investigated, resolved, and the resolution documented in the report.

# Acknowledgements

No acknowledgments to declare.

# Disclosures

JNK declares consulting services for Bioptimus, France; Owkin, France; DoMore Diagnostics, Norway; Panakeia, UK; AstraZeneca, UK; Scailyte, Switzerland; Mindpeak, Germany; and MultiplexDx, Slovakia. Furthermore he holds shares in StratifAI GmbH, Germany, has received a research grant by GSK, and has received honoraria by AstraZeneca, Bayer, Daiichi Sankyo, Eisai, Janssen, MSD, BMS, Roche, Pfizer and Fresenius. DT received honoraria for lectures by Bayer and holds shares in StratifAI GmbH, Germany. SF has received honoraria from MSD and BMS. RL declares consulting services and honoraria from MSD, Janssen, AstraZeneca, Astellas, Roche. AM has received honoraria as a consultant, advisor or speaker from Roche, Lilly and Menarini/Stemline, and has received support for accommodation and travel from AstraZeneca, all outside the submitted work. OSMEN holds shares in StratifAI GmbH, Germany. No other COIs are declared by any of the authors.

# Funding

JNK is supported by the German Cancer Aid (DECADE, 70115166), the German Federal Ministry of Education and Research (PEARL, 01KD2104C; CAMINO, 01EO2101; SWAG, 01KD2215A; TRANSFORM LIVER, 031L0312A; TANGERINE, 01KT2302 through ERA-NET Transcan; Come2Data, 16DKZ2044A; DEEP-HCC, 031L0315A), the German Academic Exchange Service (SECAI, 57616814), the German Federal Joint Committee (TransplantKI, 01VSF21048) the European Union's Horizon Europe and innovation programme (ODELIA, 101057091; GENIAL, 101096312), the European Research Council (ERC; NADIR, 101114631), the National Institutes of Health (EPICO, R01 CA263318) and the National Institute for Health and Care Research (NIHR, NIHR203331) Leeds Biomedical Research Centre. The views expressed are those of the author(s) and not necessarily those of the NHS, the NIHR or the Department of Health and Social Care. This work was funded by the European Union. Views and opinions expressed are however those of the author(s) only and do not necessarily reflect those of the European Union. Neither the European Union nor the granting authority can be held responsible for them. SF is supported by the German Federal Ministry of Education and Research (SWAG, 01KD2215C), the German Cancer Aid (DECADE, 70115166 and TargHet, 70115995) and the German Research Foundation (504101714). The DACHS study was supported by the German Research Council (BR 1704/6-1, BR 1704/6-3, BR 1704/6-4, CH 117/1-1, HO 5117/2-1, HO 5117/2-2, HE 5998/2-1, HE 5998/2-2, KL 2354/3-1, KL 2354 3-2, RO 2270/8-1, RO 2270/8-2, BR 1704/17-1, BR 1704/17-2); the Interdisciplinary Research Program of the National Center for Tumor Diseases (NCT), Germany; and the German Federal Ministry of Education and Research (01KH0404, 01ER0814, 01ER0815, 01ER1505A, and 01ER1505B). AM is supported by the European Society for Medical Oncology José Baselga Fellowship for Clinician Scientists founded by AstraZeneca (2023–2025).

# Figures

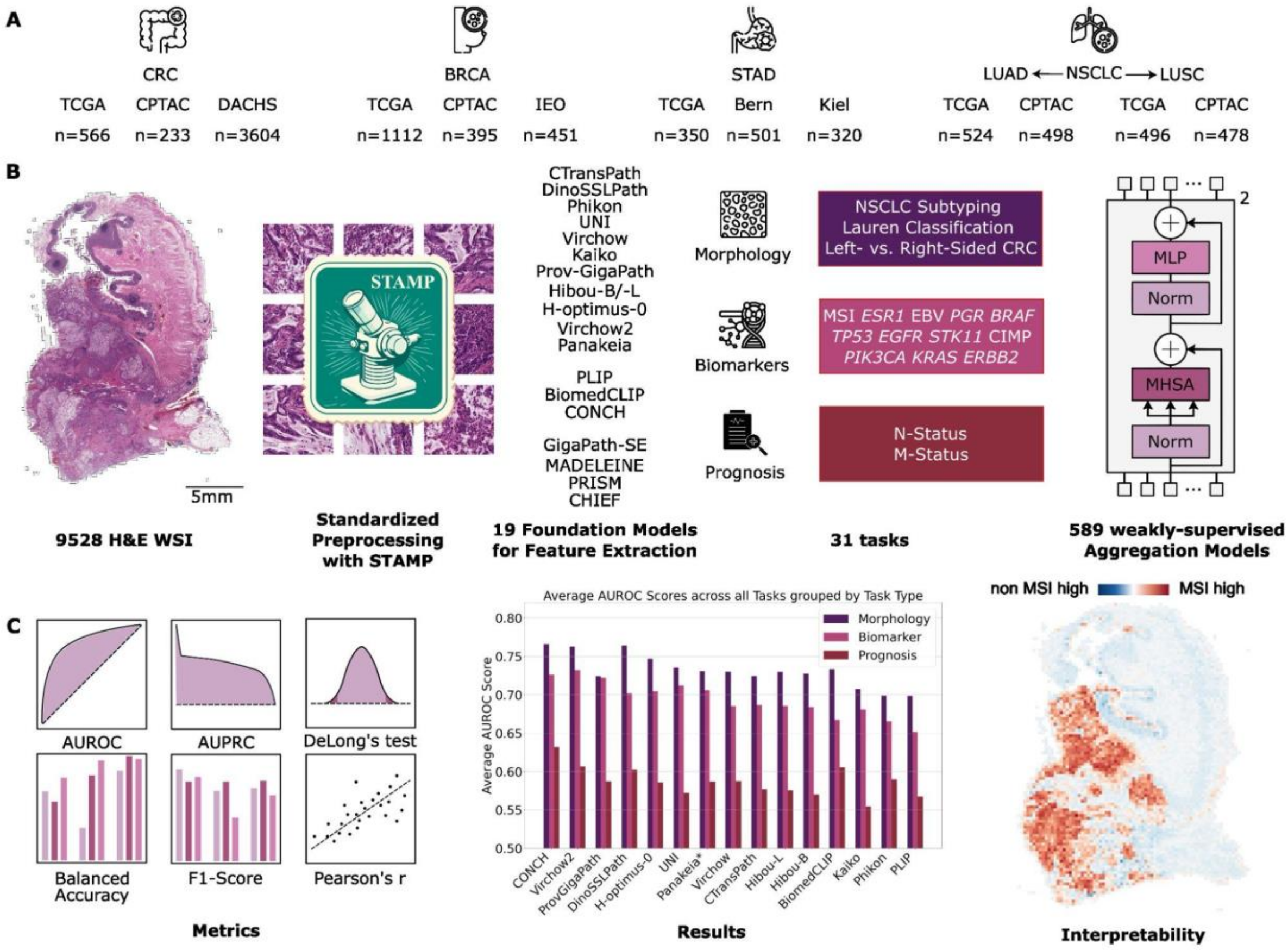

**Fig. 1: Experimental design of the study**. Benchmarking of 19 histopathology foundation models using 13 cohorts and 31 tasks. **A**, number of slides used from each of the 13 cohorts including four cancer types. **B**, 9528 hematoxylin and eosin (H&E) stained whole slide images (WSIs) were preprocessed using the standardized STAMP[19] pipeline. Feature extraction from the processed tiles was performed using 19 foundation models analyzed in this study. The TCGA features were utilized for five-fold cross-validation with downstream Transformer models on 31 classification tasks using STAMP. All models were subsequently applied to external features from CPTAC, Bern, Kiel, DACHS, and IEO. **C**, All experiments were analyzed using AUROCs, supplemented by AUPRC, Pearson's correlation coefficient, DeLong's test, balanced accuracy and F1-score. CONCH achieves the highest average AUROC across all tasks, followed by Virchow2, Prov-GigaPath and DinoSSLPath. The star indicates Panakeia was tested on all tasks despite being specifically designed for BRCA and CRC. Attention heatmaps were generated for some slides to interpret differences between foundation models.

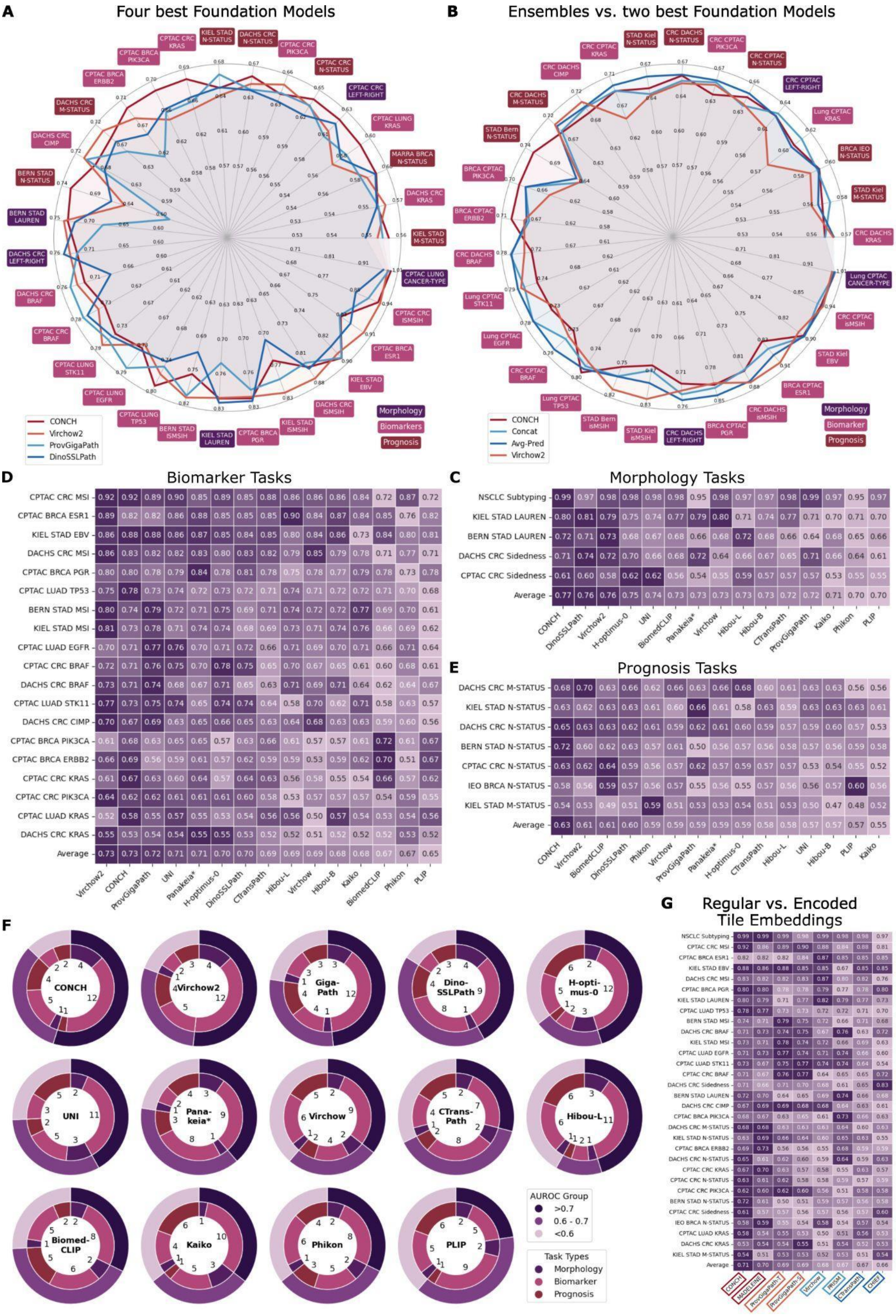
A
Four best Foundation Models
CONCH
Virchow2
ProvGigaPath
DinoSSLPath
Morphology
Biomarkers
Prognosis
B
Ensembles vs. two best Foundation Models
CONCH
Concat
Avg-Pred
Virchow2
D
Biomarker Tasks
CPTAC CRC MSI
CPTAC BRCA ESR1
KIEL STAD EBV
DACHS CRC MSI
CPTAC BRCA PGR
CPTAC LUAD TP53
BERN STAD MSI
KIEL STAD MSI
CPTAC LUAD EGFR
CPTAC CRC BRAF
DACHS CRC BRAF
CPTAC LUAD STK11
DACHS CRC CIMP
CPTAC BRCA PIK3CA
CPTAC BRCA ERBB2
CPTAC CRC KRAS
CPTAC CRC PIK3CA
CPTAC LUAD KRAS
DACHS CRC KRAS
Average
Virchow2
CONCH
ProvGigaPath
UNI
Panakeia*
H-optimus-0
DinoSSLPath
CTransPath
Hibou-L
Virchow
Hibou-B
Kaiko
BiomedCLIP
Phikon
PLIP
C
Morphology Tasks
NSCLC Subtyping
KIEL STAD LAUREN
BERN STAD LAUREN
DACHS CRC Sidedness
CPTAC CRC Sidedness
Average
CONCH
DinoSSLPath
Virchow2
H-optimus-0
UNI
BiomedCLIP
Panakeia*
Virchow
Hibou-L
Hibou-B
CTransPath
ProvGigaPath
Kaiko
Phikon
PLIP
E
Prognosis Tasks
DACHS CRC M-STATUS
KIEL STAD N-STATUS
DACHS CRC N-STATUS
BERN STAD N-STATUS
CPTAC CRC N-STATUS
IEO BRCA N-STATUS
KIEL STAD M-STATUS
Average
CONCH
Virchow2
BiomedCLIP
DinoSSLPath
Phikon
Virchow
ProvGigaPath
Panakeia*
H-optimus-0
CTransPath
Hibou-L
UNI
Hibou-B
PLIP
Kaiko
F
CONCH
Virchow2
Giga-Path
Dino-SSLPath
H-optimus-0
UNI
Panakeia*
Virchow
CTrans-Path
Hibou-L
Biomed-CLIP
Kaiko
Phikon
PLIP
AUROC Group
>0.7
0.6 - 0.7
<0.6
Task Types
Morphology
Biomarker
Prognosis
G
Regular vs. Encoded Tile Embeddings

**Fig. 2: Performance of 19 pathology foundation models on 31 weakly-supervised prediction tasks. A**, Area under the receiver operator characteristic (AUROC) scores of the four best foundation models, taskwise normalization. **B**, AUROC scores of the two best foundation models compared to the average prediction of the four best models (Avg-Pred) and the concatenated vectors of CONCH and Prov-GigaPath (Concat). **C-E**, Average AUROC scores of the five folds of each foundation model on Morphology (**C**), Biomarker (**D**) and Prognosis (**E**) tasks. Taskwise normalization for better comparison of the foundation models. Tasks are sorted by their mean AUROC across all models, while models are sorted by their mean AUROC across all tasks. **F**, stacked pie charts showing the number of tasks where each model achieved an average AUROC of >0.7, 0.6 - 0.7, or <0.6, grouped by task type. **G**, Average AUROC scores of the five folds using encoded tile embeddings from slide encoders vs. the original tile embeddings. The star indicates Panakeia was tested on all tasks despite being specifically designed for BRCA and CRC.

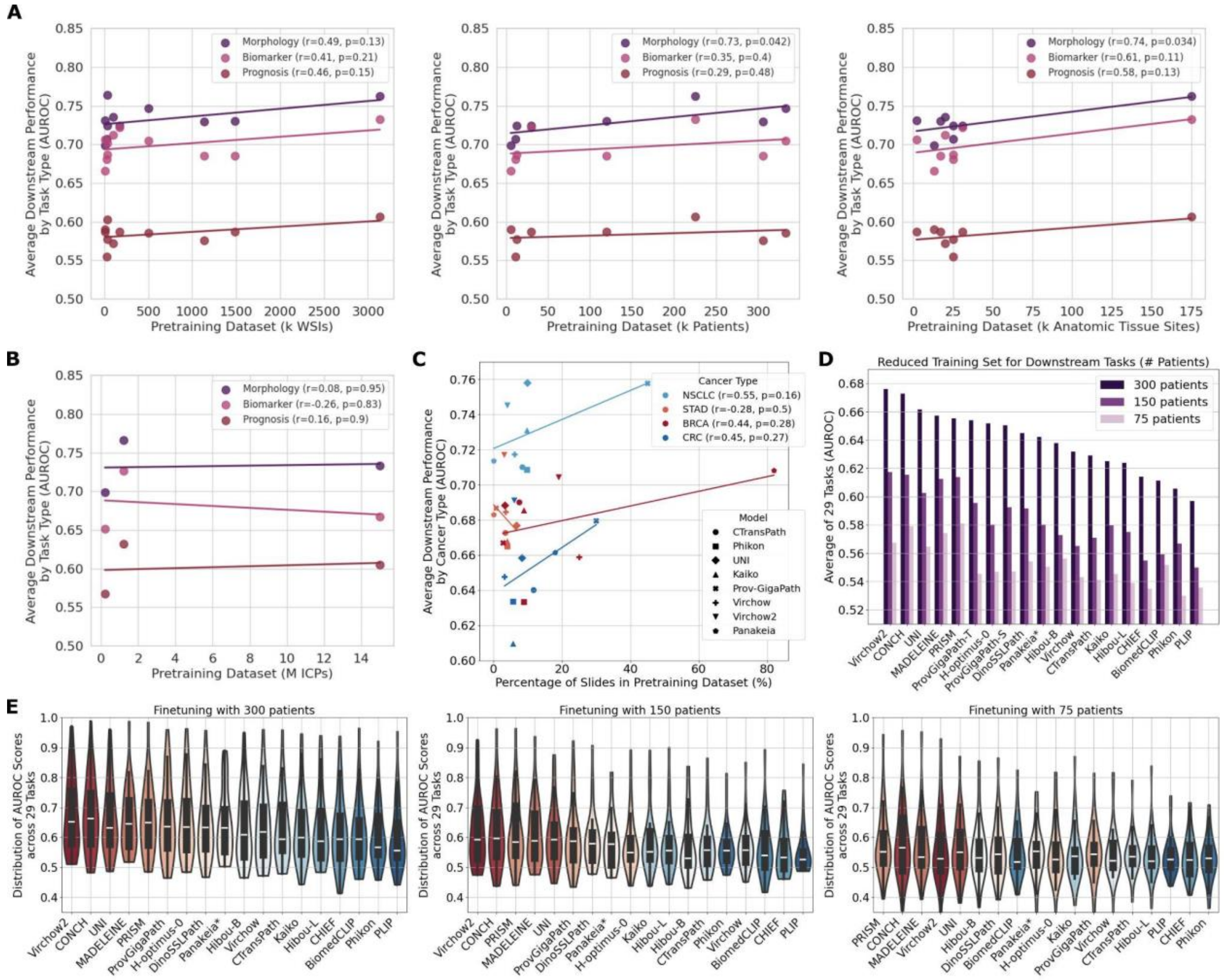

**Fig. 3: The impact of data diversity and volume on downstream weakly-supervised classification performance. A,B,C**, The impact of foundation model data diversity on downstream classification. Correlation between the number of WSIs, patients and anatomic tissue sites in the pretraining dataset and the average AUROC for each downstream task type for all vision-only foundation models for which this data is available (**A**). Correlation between the number of Image-Caption Pairs (ICPs) in the pretraining dataset and the average AUROC for each downstream task type for all vision-language foundation models (**B**). Performance of the respective cancer types correlated with the proportion of the cancer type in the pretraining dataset (**C**). All information that was available is shown (**Table S3, S4, S5**). **D-E**, Experiments with reduced downstream training sizes. Average AUROC scores across 29 tasks, trained with 75, 150, or 300 patients (**D**). Distribution of AUROC scores across all tasks for each model separately (**E**). The star indicates Panakeia was tested on all tasks despite being specifically designed for BRCA and CRC.

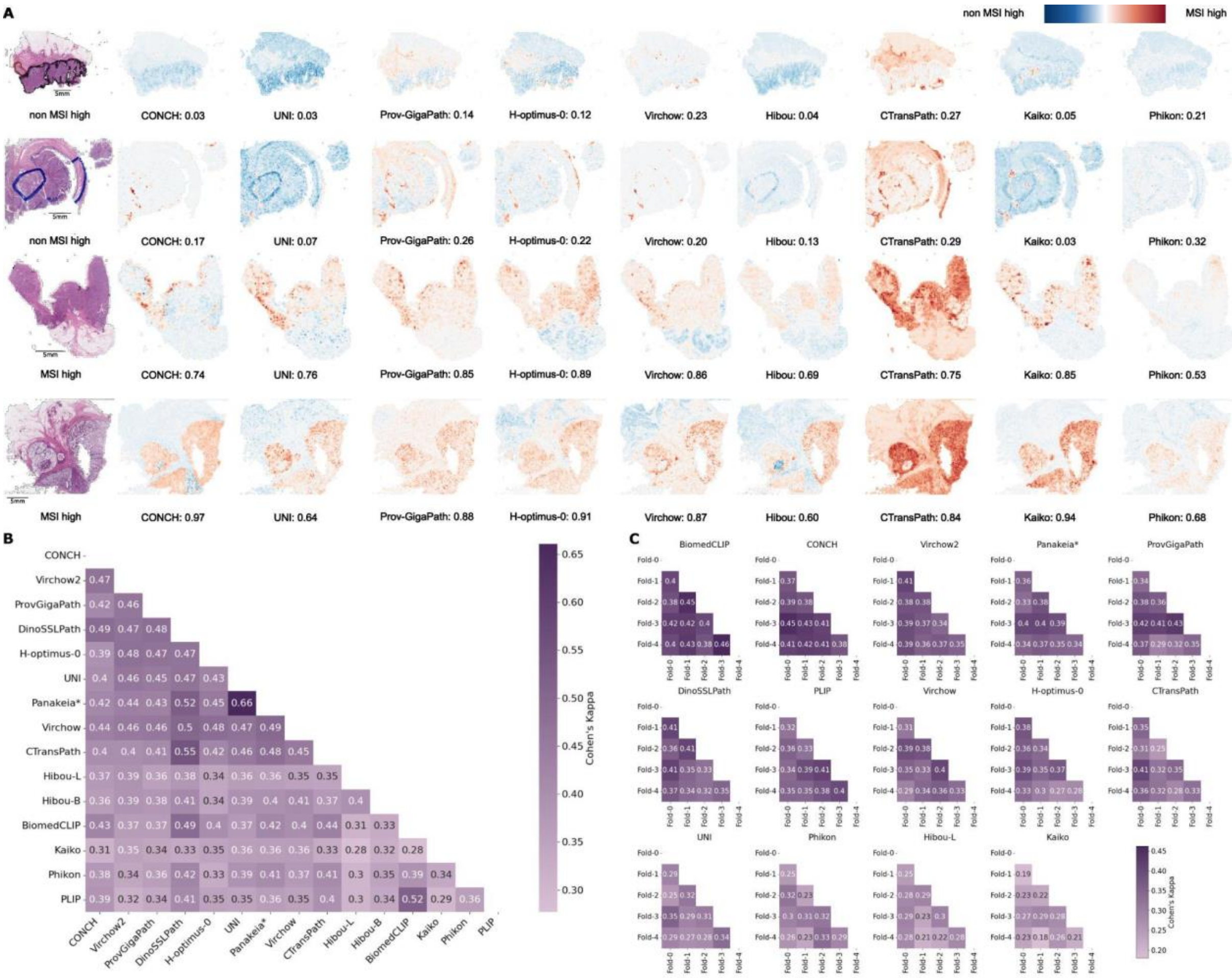

**Fig. 4: Divergence in tissue focus and predictive similarity among foundation models.** **A**, Attention Heatmap Analysis for MSI-H Classification in four different DACHS samples selected for correct predictions across selected foundation models. Thumbnails of the original whole slide images (WSIs) and heatmaps of selected foundation models. **B**, Objective measure of similarity of prediction scores using Cohen's Kappa and majority vote across the five folds to binarize the predictions. Kappa scores of all combinations of foundation models tested in this study. **C**, Cohen's Kappa between the five folds of each foundation model. The star indicates Panakeia was tested on all tasks despite being specifically designed for BRCA and CRC.

# Material and Methods

## Ethics statement

This study was carried out in accordance with the Declaration of Helsinki. The Clinical Proteomic Tumor Analysis Consortium (CPTAC) and TCGA did not require formal ethics approval for a retrospective study of anonymised samples. The analysis of the testing cohort DACHS (an epidemiological study which is led by the German Cancer Research Center, DKFZ, Heidelberg, Germany) was approved by the ethics committee of the Medical Faculty, University of Heidelberg under 310/2001 [30–32].

## Datasets

The study utilized datasets from TCGA, CPTAC and proprietary cohorts. Specifically, cohorts from lung adenocarcinoma (LUAD), lung squamous cell carcinoma (LUSC), colorectal cancer (CRC), stomach adenocarcinoma (STAD), and breast cancer (BRCA) were included. TCGA datasets were used for training of the models and CPTAC, DACHS, Kiel, Bern and IEO were used for evaluation. This ensured that all testing was done on data that had neither been seen during training of the foundation models nor the aggregator models. For our analyses, we only use the CPTAC-2 and CPTAC-3 prospective collections (from 2018/20), which exclusively contain patients with CPTAC-IDs and have no overlap with TCGA patients.

For external validation, CPTAC datasets for LUAD, LUSC, colorectal adenocarcinoma (COAD), and BRCA were used. No foundation models analyzed in this study were trained on CPTAC, ensuring its suitability as an independent test cohort. Additionally, for CRC, the DACHS cohort was utilized alongside CPTAC as another external test set. In STAD, proprietary datasets from Kiel and Bern served as external validation cohorts. For BRCA, the IEO dataset was used alongside CPTAC for external validation (**Figure 1A, S12**).

## Experimental Design

DP involves several task categories, including morphological, biomarker and prognostic tasks, and foundation models should be capable of performing well across all of them. In this study, we assembled and benchmarked 19 foundation models - the 12 pure vision models CTransPath, DinoSSLPath, Phikon, UNI, Virchow, Kaiko (ViT-L/14), Prov-GigaPath, Hibou-B, Hibou-L, H-optimus-0, Virchow2 and Panakeia, the 3 vision-language models PLIP, BiomedCLIP and CONCH, and the 4 Slide Encoders GigaPath, MADELEINE, PRISM and CHIEF - across a comprehensive set of tasks from all three categories. Each category was assessed across all cancer types, apart from morphological features in BRCA and prognostic features in NSCLC due to data unavailability. Biomarkers were selected based on clinical relevance, diversity, and availability. Tasks were prioritized when they were associated with actionable therapeutic targets, as annotated by OncoKB[33]. To enable both training and independent testing, each task required ground truth data to be available in TCGA (for training) and at least one test cohort. For each cohort, only tasks with at least 10 cases in each category were included (**Table S6**). For visualization purposes, only 15 models (vision-only and vision-language models) are displayed in most figures. The slide encoders were included selectively, such as in **Figure 2G** for comparison with their tile embedding counterparts and in **Figures**

**3D, 3E, S6** to highlight their potential benefits in scarce data settings. **Figures S1, S10** include all models to comprehensively display all experiments.

First, we investigated morphological classification tasks related to cancer subgroups with distinct phenotypic characteristics. The aim was to assess foundation models by evaluating their ability to discern established phenotypic distinctions. In CRC, the morphological task involved predicting whether the slide originated from the left or right side of the colon, excluding colon transversum samples due to ambiguous classification. In STAD, the Lauren classification [34] was chosen as the morphological task, classifying slides as “intestinal”, “diffuse”, or “mixed”, given the unavailability of ground truth for newer classification systems [35,36]. In lung cancer, the models were tasked with classifying samples into either adenocarcinoma or squamous cell carcinoma [1].

Biomarker prediction tasks focused mainly on clinically relevant targets with some type of morphological correlation as demonstrated by previous computational pathology models. For CRC, these included *BRAF*, *KRAS*, microsatellite instability (MSI) status, *PIK3CA*, and CpG island methylator phenotype (CIMP) status [11]. For STAD, Epstein-Barr virus (EBV) presence and MSI status were selected [37]. For LUAD, the targets were *EGFR*, *STK11*, *KRAS*, and *TP53* [1]. For BRCA, the targets were the expression of HER2, ER, PR receptors, and *PIK3CA* mutations [38,39]. MSI status and CIMP status were binarized into MSI-high versus not MSI-high and CIMP-high versus not CIMP-high, respectively. HER2, ER, and PR expression were binarized using the z-score of mRNA expression profiles, similar to a study by Wegscheider et al. [40]. This approach was preferred over immunohistochemistry labels due to its objectivity and reduced variance error.

Prognostic tasks, which aim to predict clinical outcomes directly from whole-slide images (WSIs), were selected based on their prognostic relevance. The tasks included N status for CRC, STAD, and BRCA, where all stages except N0 were classified as N+ (excluding Nx cases). M-Status was analyzed in CRC and STAD, performing binary classification of M0 versus M+.

By focusing on tasks with clear therapeutic actionability or prognostic relevance, we aimed to evaluate the practical utility of these models in a clinical setting. This comprehensive benchmarking study included 31 tasks across 8 external test cohorts, encompassing a wide range of clinically relevant classification tasks (**Table S7**).

# Image Processing and Deep Learning Techniques

The benchmarking was conducted using the STAMP pipeline version 1.1.0 [19] (**Table S8**). Each classification task followed a two-step procedure (**Figure 1B**). In the first step, feature vectors were extracted from WSIs utilizing the foundational models evaluated in this study. In the second step, these vectors were employed to train a slide-level aggregator on the downstream tasks described above.

WSIs were segmented into N tiles, with an edge length of 224 px corresponding to 256 µm, resulting in an effective resolution of ~1.14 µm per pixel. All included foundation models in our benchmark, except for Prov-GigaPath[22], tessellate the slide into tiles of 224x224 pixels. However, the Prov-GigaPath implementation transforms tiles using center cropping from 256x256 into 224x224 before inputting it into the tile encoder. The slide encoder then processes these feature embeddings generated by the tile encoder, implicitly maintaining the 224×224 tile dimensionality throughout the pipeline. Therefore, our choice of tile dimensionality for slide tessellation is consistent with the foundation models selected for our analyses. Background tiles were excluded using Canny edge detection [41]. Feature extraction was performed on each tile individually using the different foundational models. The embedding dimensions M varied across models, ranging from M=384 for DinoSSLPath and Panakeia to M=1536 for Prov-GigaPath and H-optimus-0. Subsequently, each slide was transformed into a 2D matrix with dimension NxM. The extracted feature vectors were input into a Transformer-based aggregator model [4]. It utilizes multi-head attention, Gaussian Error Linear Unit (GELU) activation functions [42], layer normalization, and a multilayer perceptron (MLP) head to produce an output corresponding to the k possible classes for each task. A five-fold cross-validation approach was implemented, resulting in the creation of 2,945 models (19 foundation models, 31 tasks, 5 folds) trained exclusively on TCGA datasets. We implemented stratified k-fold cross-validation to ensure each fold maintains representative proportions of all classes, preventing scenarios where rare categories have zero instances in training runs. This approach follows standard practices in computational pathology and provides robust performance estimates and better generalization assessment [10]. All experiments were run on individual 40 GB NVIDIA RTX A6000 and L40 GPU (graphics processing unit) nodes. In addition to the transformer-based aggregator described, we evaluated ABMIL as an alternative aggregation method [24]. ABMIL introduces inductive bias by using attention mechanisms to assign weights to each tile in a slide, enabling the model to focus on the most informative regions.

To integrate slide encoders into the MIL pipeline, we extracted the encoded tile-level embeddings for Prov-GigaPath, MADELEINE, CHIEF, and the 512 latents for PRISM. These encoded tile embeddings were subsequently treated as regular tile embeddings in all analyses. Unless explicitly stated otherwise, results presented throughout the study refer to the regular tile embeddings. Prov-GigaPath provides both a slide-level and a tile-level encoder and we evaluated both approaches [22]. In the case of Virchow and Virchow2, Vorontsov et al. proposed concatenating the class token with the average pool of patch tokens for each tile embedding. To maintain consistency with other models that only use class tokens, two configurations were tested: one including and one excluding the averaged patch tokens. As the differences are very small, the version only using class tokens is shown in the main results for consistency with other models. For CONCH, we used the output of the attentional pooler that corresponds to image-text alignment, with an embedding dimension of 512. Although the Panakeia models are specifically designed for BRCA and CRC, respectively, we also evaluate the CRC model on STAD and the BRCA model on NSCLC. This is because their performance remains competitive in these contexts, and including these results provides the basis for comparison in subsequent analyses. For experiments involving combined feature vectors, vectors were concatenated, maintaining a single vector per tile. For instance, combining CONCH and Virchow2 resulted in a combined embedding dimension M of M=1,792 (M=512 for CONCH + M=1,280 for Virchow2).

## Explainability

To better interpret the output of the models, we generated whole-slide prediction heatmaps for selected tasks. These heatmaps illustrate the models' focus on specific tissue areas, by weighting the scores assigned to individual tiles using Gradient-weighted Class Activation Mapping (Grad-CAM) [43]. It is important to note that a high number of positively contributing tiles do not automatically result in a high final score due to the non-linear aggregation process in neural networks [44]. The benchmarking effort involved 2,945 models and 9,528 slides, leading to a vast number of model-slide combinations. Thus, it was necessary to select a few informative examples methodically. Slides were selected by including cases where models showed strong disagreements and cases where all models performed well. The heatmaps were visually analyzed and compared to the underlying WSI. To further analyze the similarity between different models, Cohen's kappa[25] was measured between each pair of foundation models.

## Statistical Analysis

The performance of the models was evaluated using the Area Under the Receiver Operating Characteristic curve (AUROC) employing five-fold cross-validation and deployment on external cohorts. Mean AUROC scores from the five cross-validation models deployed on external data were used for statistical and graphical evaluations. Predictions were made per patient, and all feature matrices belonging to one patient were concatenated for use in the model. In addition to AUROC, for completeness in the supplementary material, we also calculated the Area Under the Precision-Recall Curve (AUPRC), balanced accuracy, and F1 scores. The two-sided DeLong's test was used to test for statistically significant differences in AUROC scores. As the DeLong's test is only applicable when a single prediction score is available for each model and sample, the average prediction score across all five folds was employed. Due to its multi-class nature, we excluded Lauren classification tasks from this analysis. This differs from the main metrics, where the AUROC/AUPRC/F1/balanced accuracy scores represent the mean across the five folds.

## Data availability

The slides for TCGA are available at https://portal.gdc.cancer.gov/. The slides for CPTAC are available at https://proteomics.cancer.gov/data-portal. The molecular data for TCGA and CPTAC are available at https://www.cbioportal.org/. The slides and biomarker data for DACHS were generated for prior studies[45–47] with restricted access. Biomarker data for DACHS are available by requesting Authorized Access to the phs001078 study [https://www.ncbi.nlm.nih.gov/projects/gap/cgi-bin/study.cgi?study_id=phs001113.v1.p1].
Applications for access to DACHS biomarker data are reserved for Senior Investigators and NIH Investigators as defined in https://dbgap.ncbi.nlm.nih.gov/aa/wga.cgi, and upon successful application grants access to the data for 1 year with the option to renew access. The slides for DACHS can only be requested directly through the DACHS principal investigators. The contact details are listed at http://dachs.dkfz.org/dachs/kontakt.html. All other cohorts can be requested from the respective study investigators. The data generated in this study for the creation of the figures are provided in the Source Data file. Source data are provided with this paper.

## Code availability

The benchmarking experiments were built upon the open-source STAMP software. All public models tested in this study are available via GitHub (https://github.com/KatherLab/STAMP-Benchmark).

# Supplementary Figures

## Fig. S1: AUROCs, AUPRCs, balanced accuracy and F1-scores for all main experiments

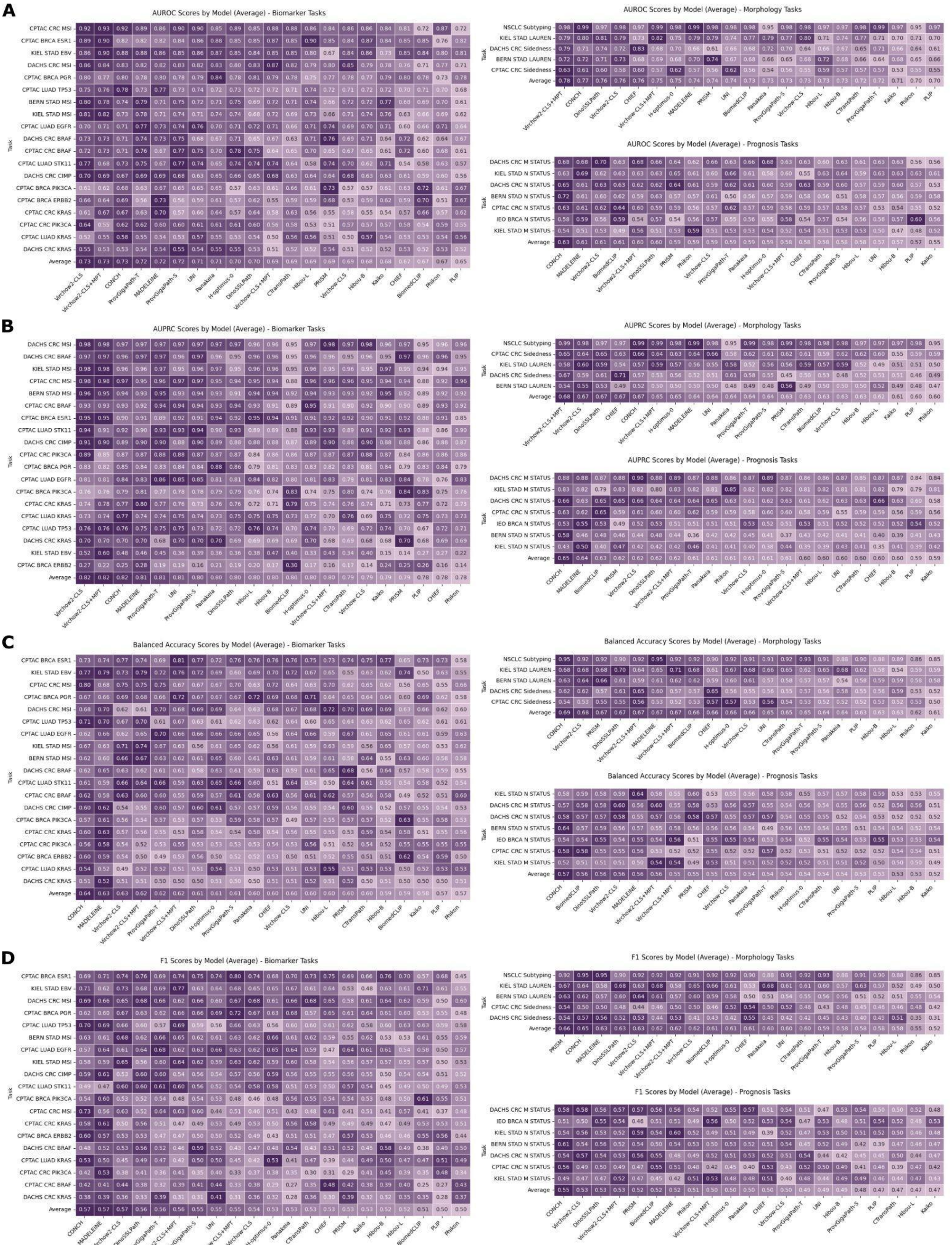

**A-D**, Average AUROC (**A**), AUPRC (**B**), balanced accuracy (**C**) and F1 (**D**) scores of the five-folds of each foundation model on Morphology, Biomarker and Prognosis tasks.

# Fig. S2: Average AUROCs sorted by cancer type and on scarce data tasks

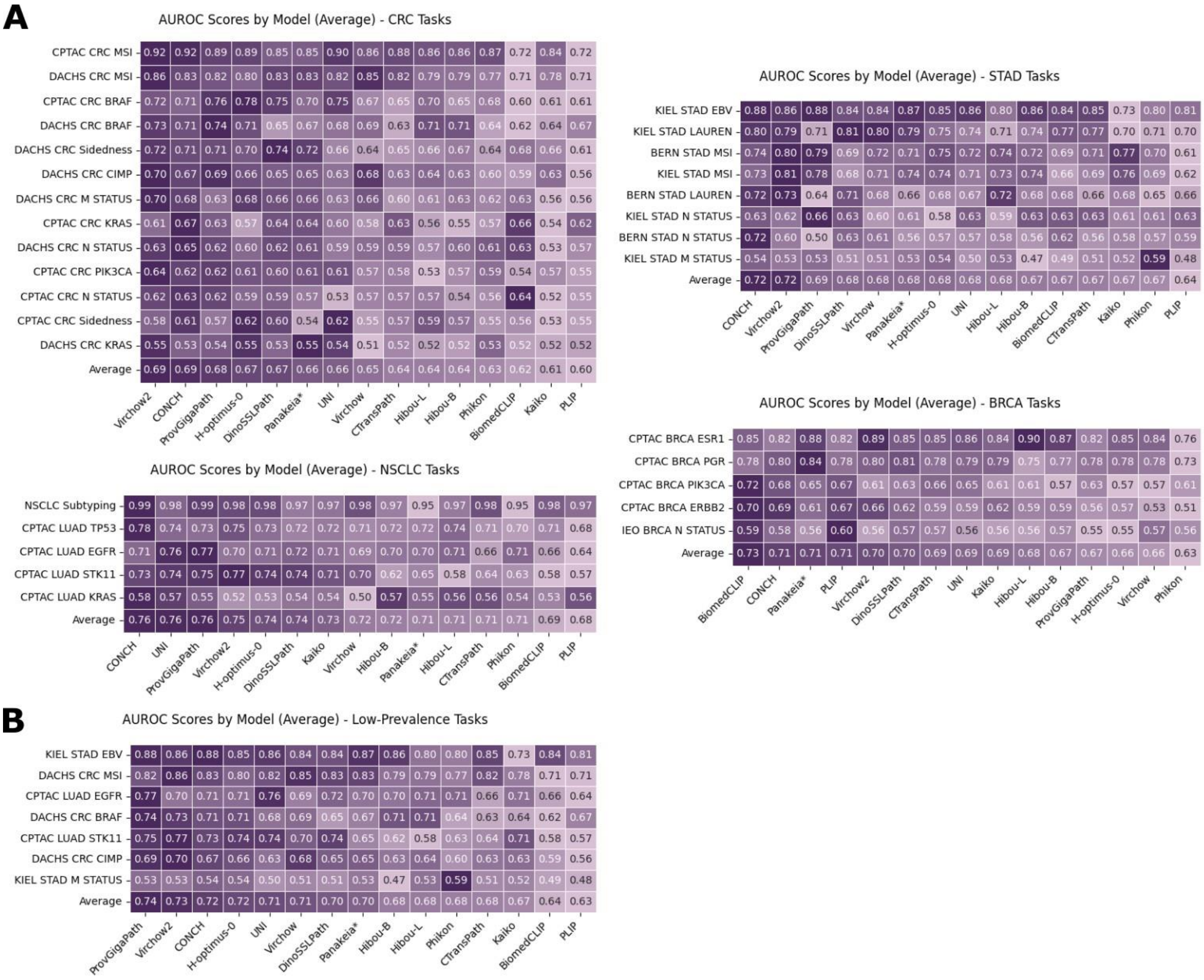

Average AUROC scores of the five folds of each foundation model. Taskwise normalization for better comparison of the foundation models. Tasks are sorted by their mean AUROC across all models, while models are sorted by their mean AUROC across all tasks. **A**, The 31 tasks were grouped by cancer type (5 tasks for NSCLC, 5 tasks for BRCA, 8 tasks for STAD, 13 tasks for CRC). Models are sorted by average performance. **B**, Only tasks with rare positive cases (>15%) in the TCGA training cohort are shown. To avoid cancer type imbalance, these tasks are only evaluated in DACHS, Kiel and CPTAC LUAD.

# Fig. S3: Comparison of STAMP and ABMIL

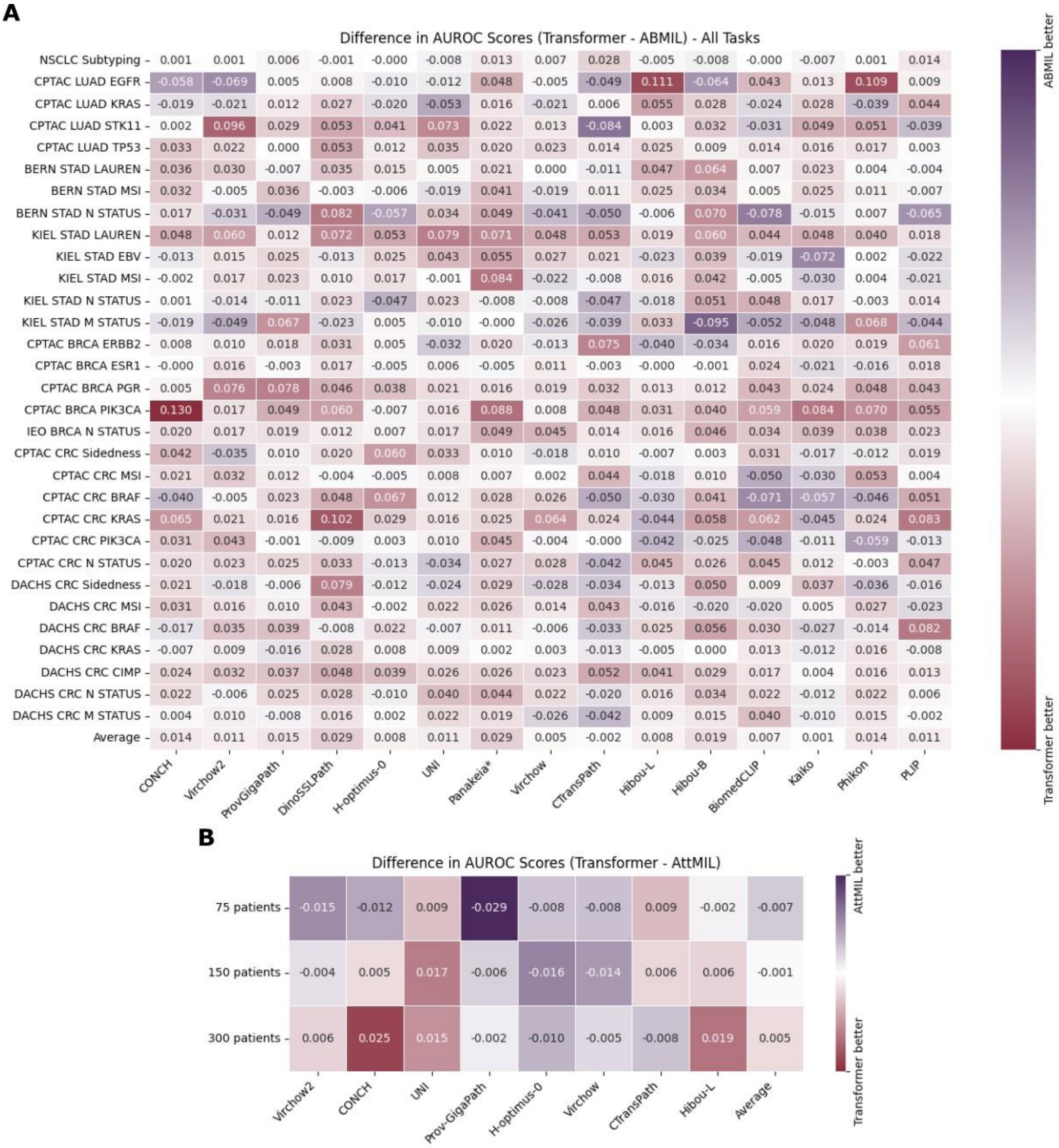

**A**, Difference in average AUROC scores between STAMP transformer-based aggregation and ABMIL across all tasks, calculated as the average over five cross-validation folds for each foundation model. Positive values indicate superior performance of STAMP. **B**, Difference in average macro-AUC scores between STAMP and ABMIL for selected foundation models under reduced downstream training dataset conditions, as shown in Fig. S6. This compares the relative performance of both methods in low-data scenarios.

# Fig. S4: Performance Comparison of Model Ensembles and Single-Model Baselines Using DeLong's Test

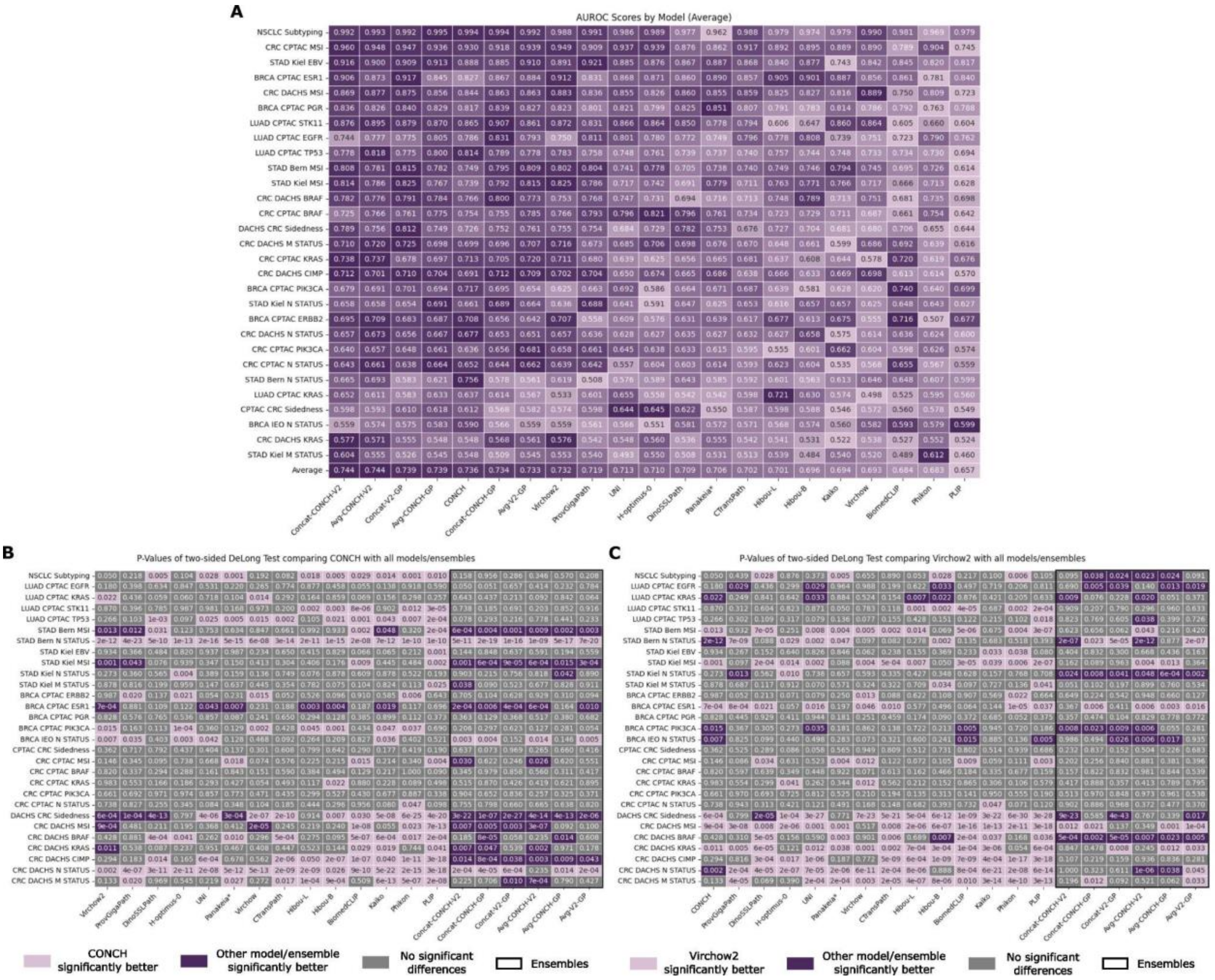

**A**, AUROC scores for each model and ensemble approach are shown, averaging predictions across five folds for individual models and five or ten folds for ensembles. Two ensembling approaches were used: taking the average prediction scores of downstream models trained on different foundation model backbones (prefix Avg) and concatenating feature vectors from different backbones to create a single downstream model (prefix Concat). The "Lauren" task was excluded as it's not a binary classification.

**B-C**, P-values from two-sided DeLong's tests comparing CONCH (**B**) or Virchow2 (**C**) with other models and ensembles. Pink indicates CONCH/Virchow2 performed significantly better, purple indicates the other model or ensemble performed significantly better, and gray shows no significant differences. No correction for multiple testing was applied; alpha was set to 0.05.

## Fig. S5: Diversity of pretraining datasets

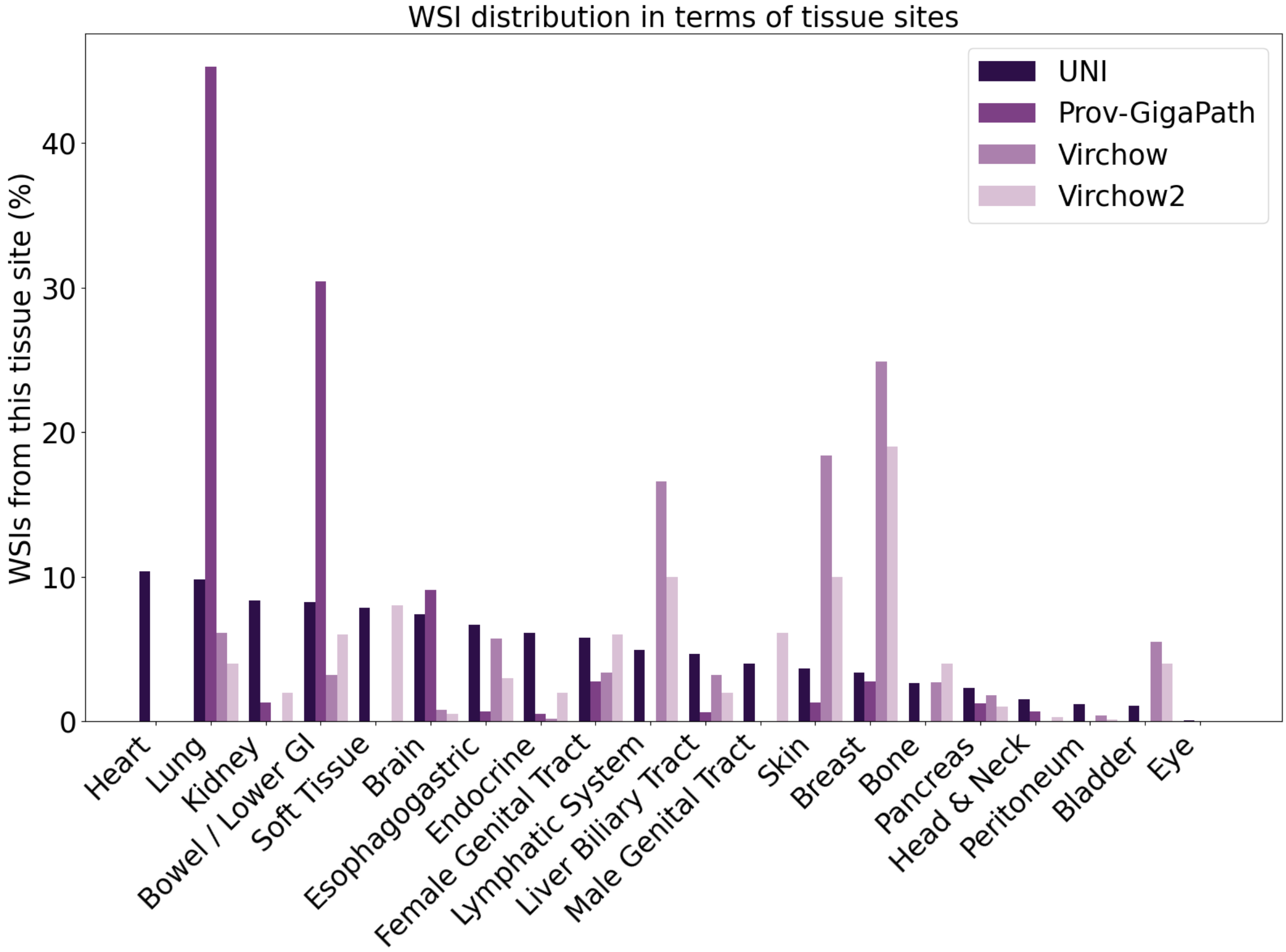

Relative number of slides per anatomic tissue site in the pretraining datasets of four foundation models. For Virchow and Virchow2, upper GI and stomach were grouped into the esophagogastric category, while cervix (Virchow2 only), endometrium, and ovary were combined into the female genital tract. Prostate and testis were combined into the male genital tract for Virchow2. Omentum was considered part of Peritoneum, Bone Marrow under Bone, Thyroid under Endocrine. For Prov-GigaPath, ovary/fallopian tube and uterus were grouped as the female genital tract, and liver with the biliary tract. Data adjustments resulted in Virchow displaying 17 tissue types in 15 categories, Prov-GigaPath with 15 tissue types in 13 categories, and Virchow2 with >175 tissue types in 18 categories.

# Fig. S6: Model performance with reduced downstream training dataset

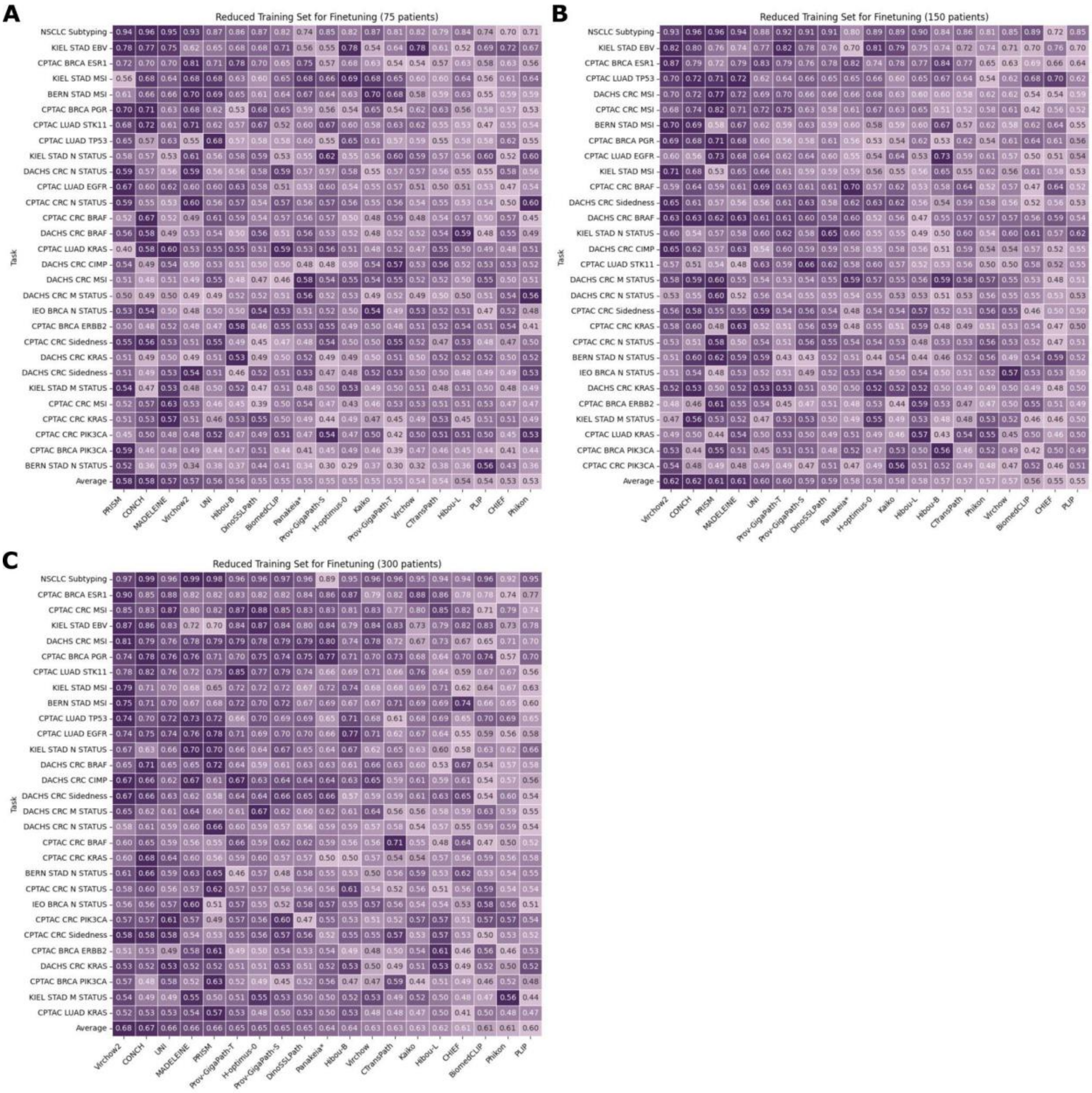

Mean AUROC across all five folds on 29 tasks for all foundation models trained with a reduced downstream dataset of 75 (**A**), 150 (**B**), or 300 patients (**C**). Patients were randomly selected from the TCGA cohorts, ensuring the ground truth was defined for all analyzed tasks. The tasks Lauren in Kiel and Bern were excluded due to insufficient patient numbers.

# Fig. S7: High-performance vs. low-performance tasks

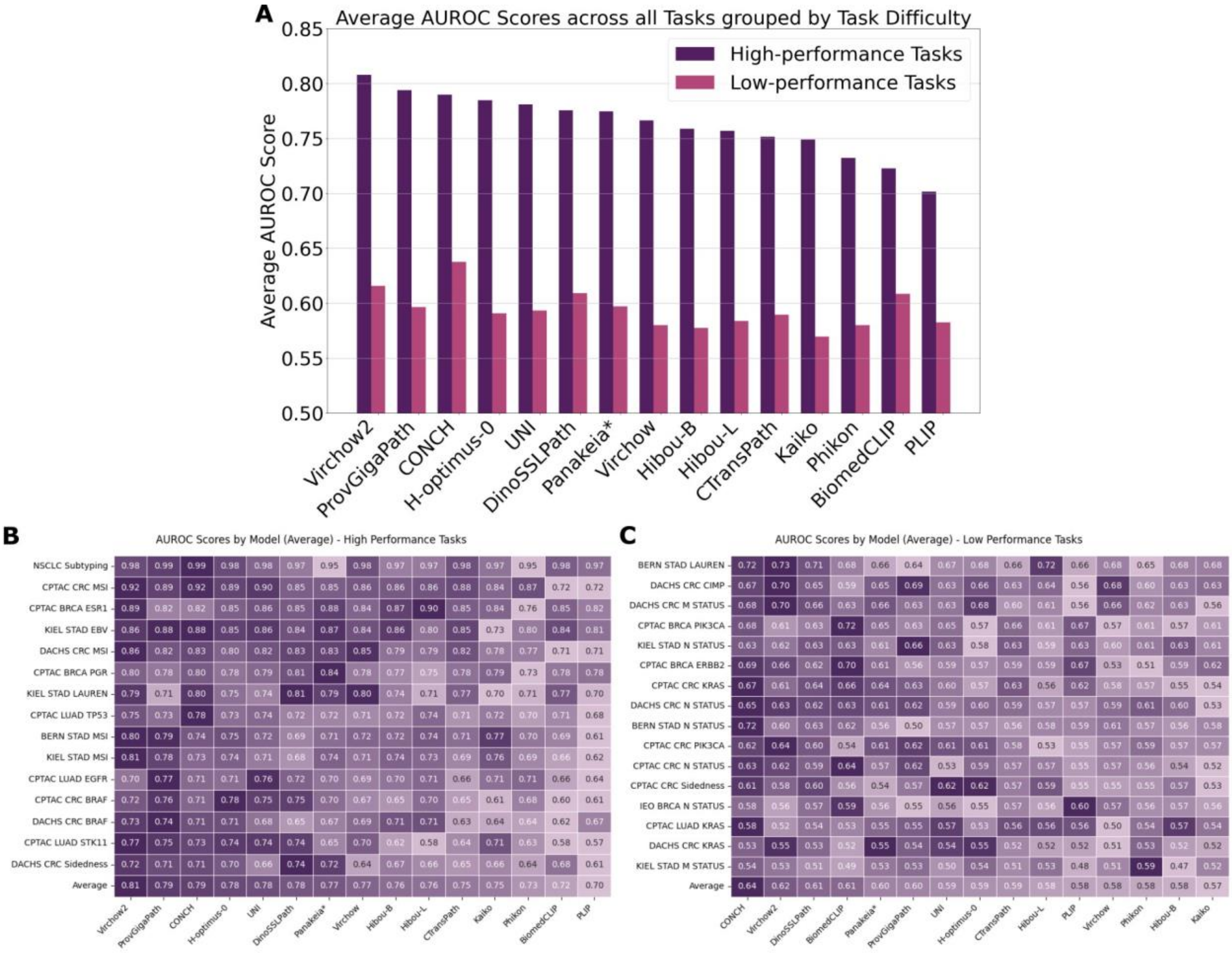

**A**, Average AUROC scores across 15 high-performance and 16 low-performance tasks. Tasks were selected by including only those where at least one foundation model achieved an average AUROC over 0.75 and all others in low-performance tasks. **B-C**, The performance of each foundation model is listed. The final row presents the overall average AUROC for each model. Tasks are sorted by their mean AUROC across all models, while models are sorted by their mean AUROC across all tasks.

# Fig. S8: Attention heatmaps of slides with large variations in prediction scores

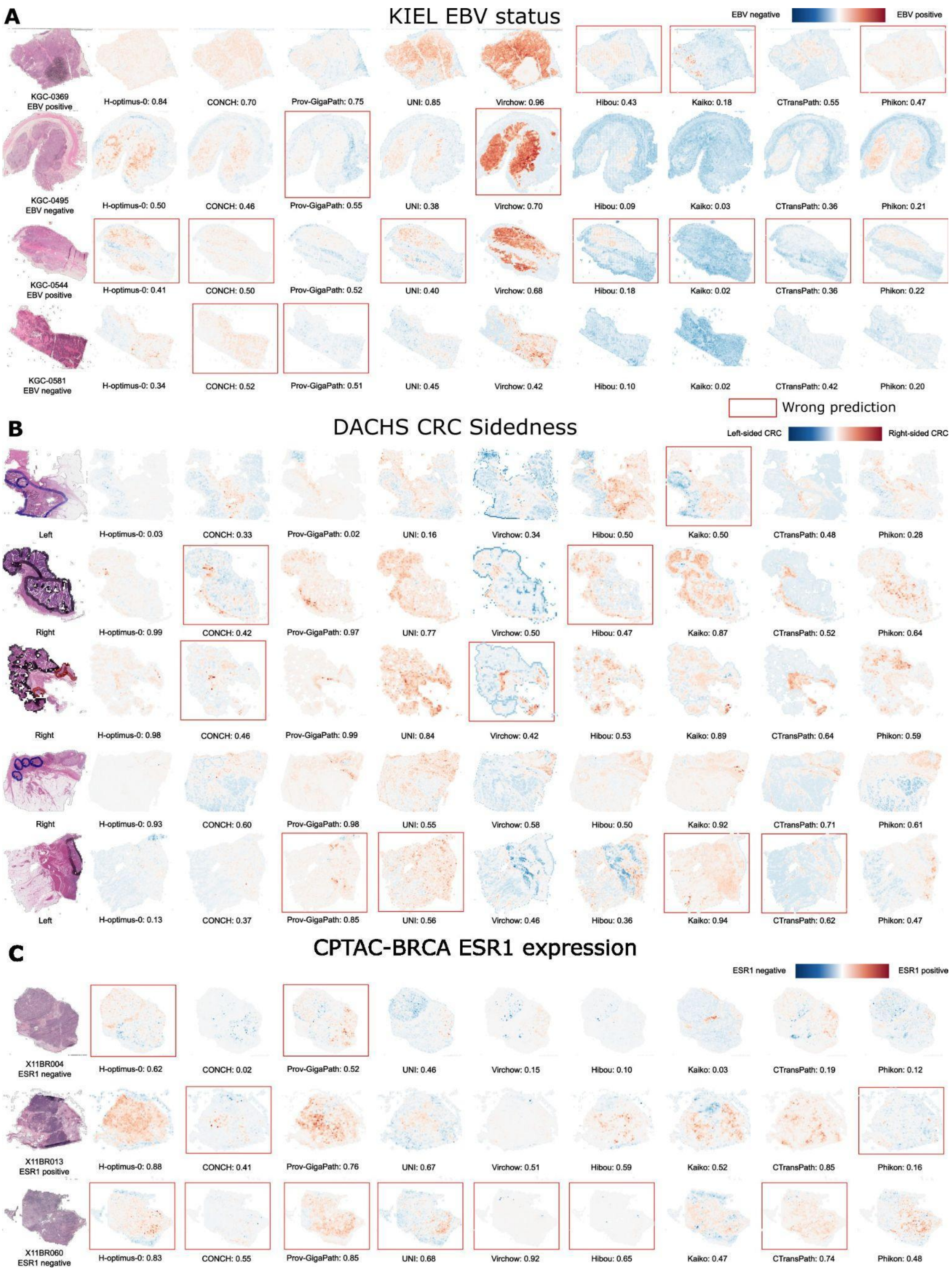

**A-C**, Attention Heatmap Analysis for Kiel EBV status (**A**), DACHS CRC sidedness (**B**) and CPTAC-BRCA ESR1 expression (**C**). Classification in 12 different WSIs selected for diverse prediction scores across the foundation models. Thumbnails of the original whole slide images (WSIs) and heatmaps of all foundation models are shown.

## Fig. S9: Attention heatmaps of slides that all models predicted well

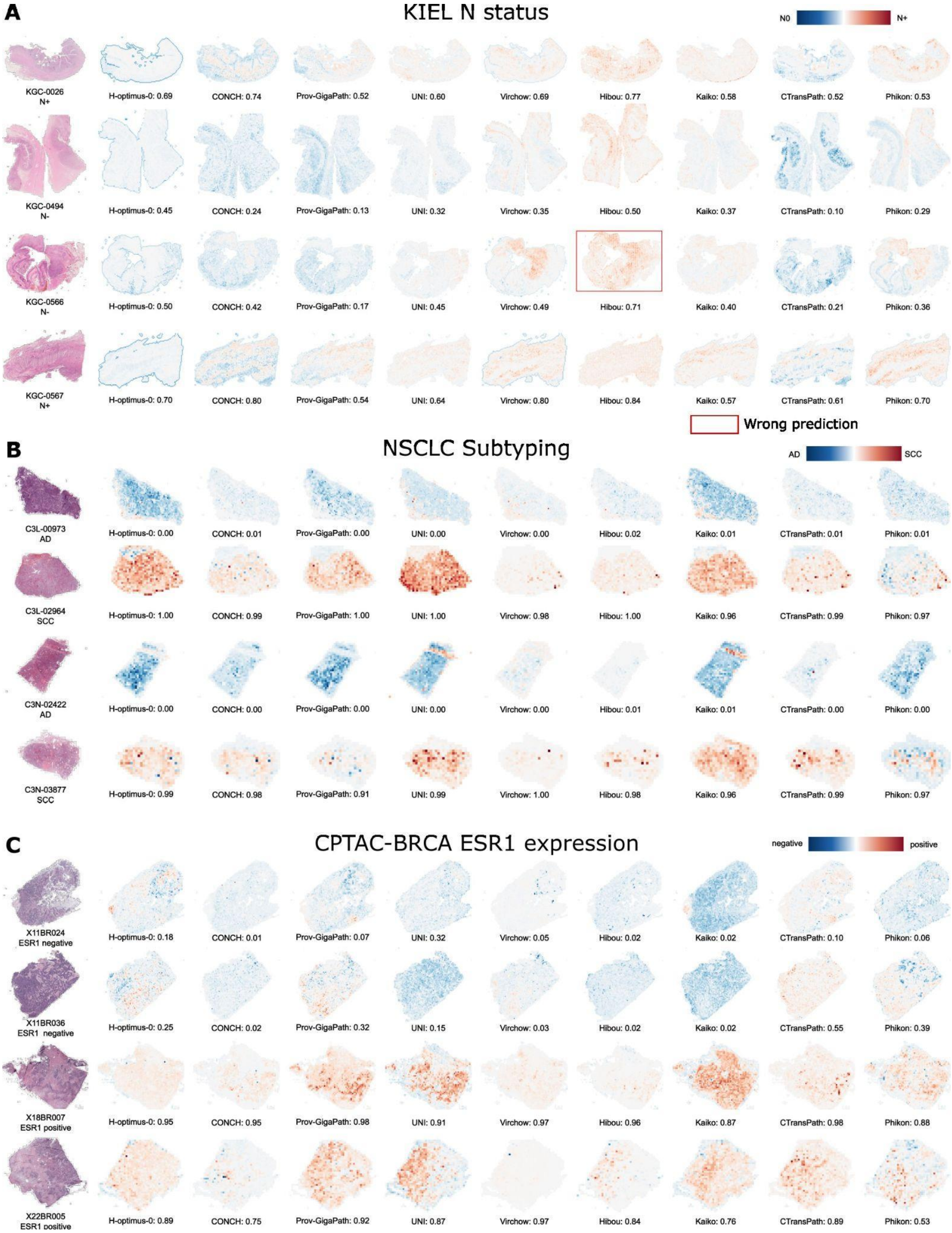

**A-C**, Attention Heatmap Analysis for Kiel N status (**A**), NSCLC subtyping (**B**) and CPTAC-BRCA ESR1 expression (**C**). Classification in four different samples per cohort selected for correct predictions across almost all foundation models. Thumbnails of the original whole slide images (WSIs) and heatmaps of all foundation models are shown.

# Fig. S10: AUROC scores across all foundation models and ensembles

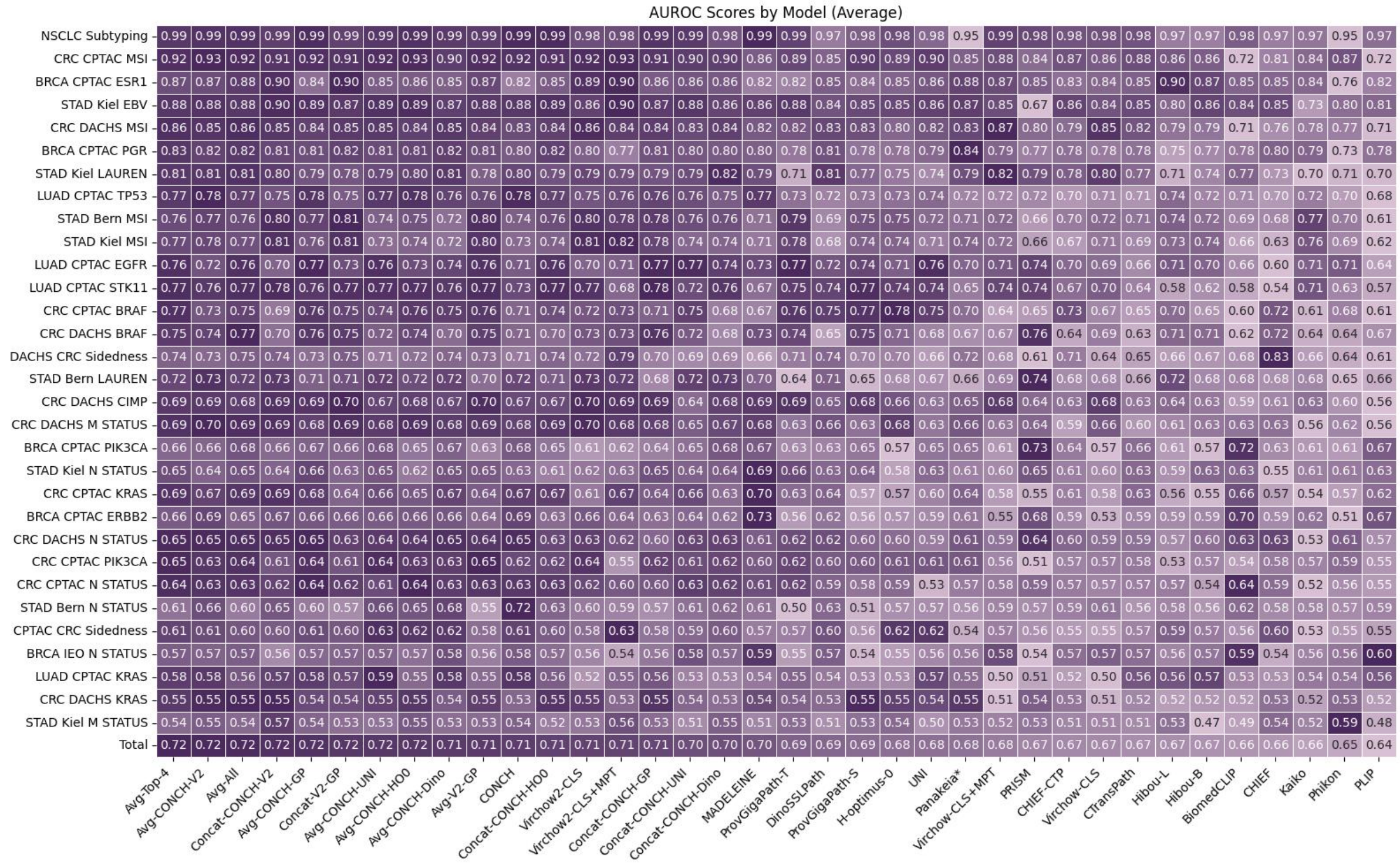

AUROC scores for all foundation models, foundation model variations, and multiple ensemble approaches. Prov-GigaPath-T are the regular tile embeddings, Prov-GigaPath-S are the tile embeddings encoded by the GigaPath slide encoder. Virchow(-2)-CLS contained only class tokens, with Virchow(-2)-CLS+MPT representing the version with class and mean patch tokens combined.

Multiple experiments were conducted using concatenated feature vectors combining features from CONCH, Virchow2, Prov-GigaPath, DinoSSLPath, H-optimus-0 and UNI.

For the same combinations, average prediction scores were calculated. These scores were used to evaluate the performance of combined predictions.

## Fig. S11: Cohen's kappa scores across all ensembles and their individual model components

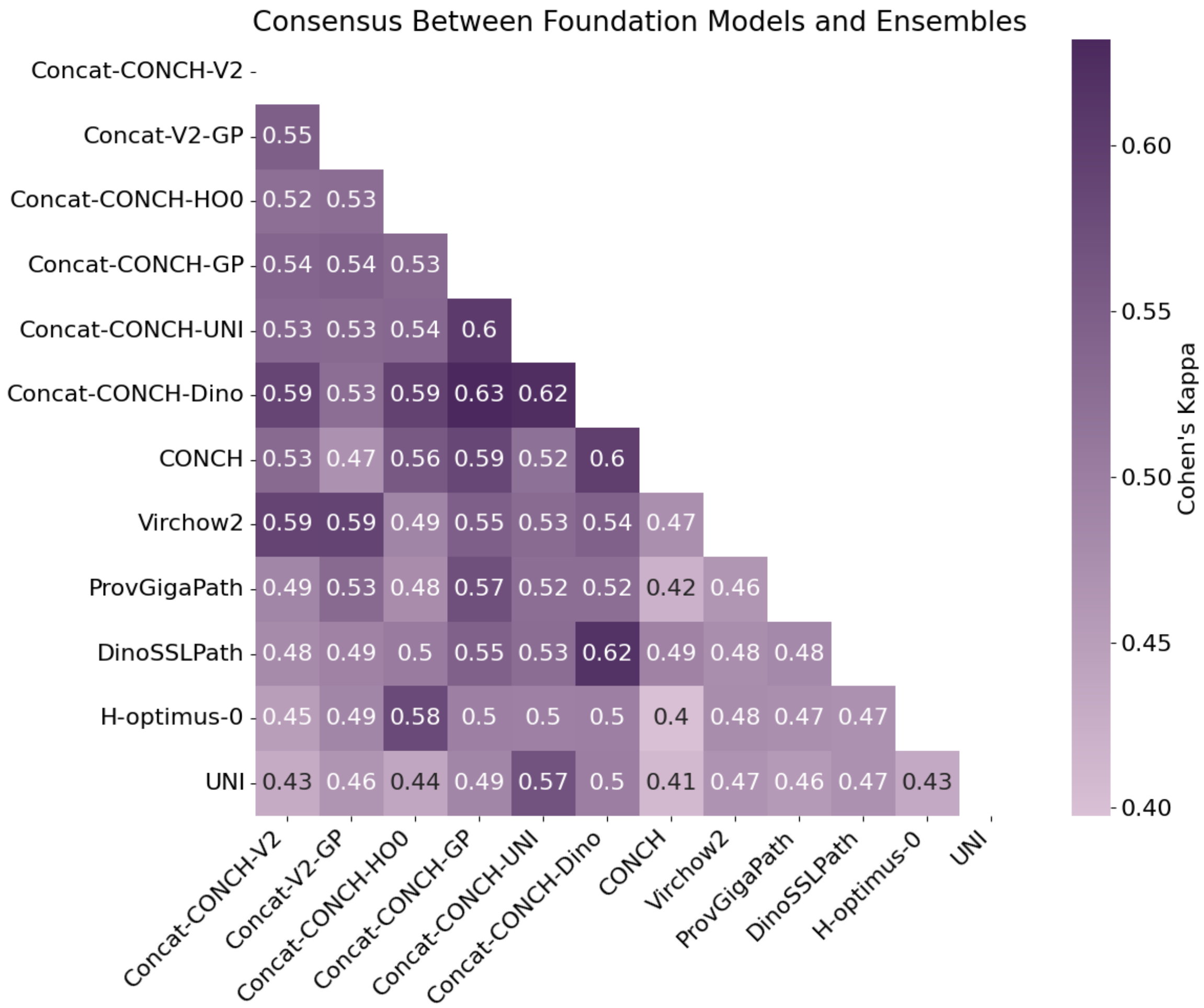

Objective measure of similarity of prediction scores using Cohen’s Kappa and majority vote across the five folds to binarize the predictions. The concatenated versions of CONCH, Virchow2 (V2), Prov-GigaPath (GP), H-optimus-0 (HO0), UNI and DinoSSLPath (Dino) and their single model counterparts are shown.

# Fig. S12: Datasets overview

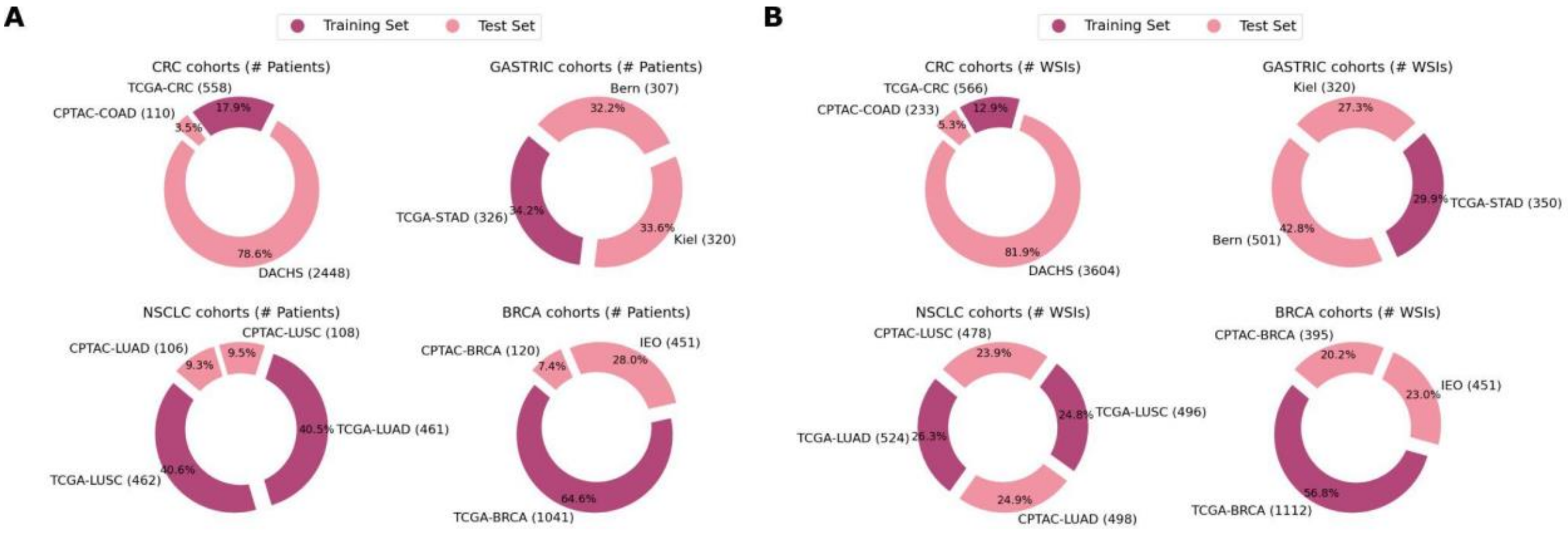

Composition of all cohorts used in this study and the size comparison of training and test sets for each cancer type. Number of patients (**A**) and number of WSIs (**B**) in each cohort. Training was conducted using the TCGA-CRC, TCGA-STAD, TCGA-LUAD, TCGA-LUSC, and TCGA-BRCA cohorts, with TCGA-BRCA being the largest training cohort. Testing was performed using the CPTAC-COAD, CPTAC-LUAD, CPTAC-LUSC, CPTAC-BRCA, DACHS, Bern, Kiel, and IEO cohorts, with DACHS being the largest testing cohort. In total, 6,818 patients and 9,528 slides were used in this study.

# Supplementary Tables

## Table S1: Analysis of Diversity Metrics in the pretraining datasets

| Model | Shannon Entropy | Simpson's Diversity Index | Evenness |
|---|---|---|---|
| UNI | 4.03 | 0.93 | 0.93 |
| Prov-GigaPath | 2.21 | 0.69 | 0.57 |
| Virchow | 3.31 | 0.86 | 0.81 |
| Virchow2 | 3.97 | 0.92 | 0.86 |

Shannon Entropy: A measure of the uncertainty or randomness in the data distribution. Higher values indicate more diversity.

Simpson's Diversity Index: Measures the probability that two individuals randomly selected from a sample will belong to different categories. Higher values indicate more diversity.

Evenness: Indicates how evenly the data are distributed across categories. It is derived from the Shannon Entropy and ranges from 0 (low evenness) to 1 (high evenness).

These analyses indicate that the UNI model has a more diversified dataset compared to Prov-GigaPath and Virchow, which might have implications for the robustness and generalizability of models trained on these datasets.

## Table S2: Accuracy and error rates of CONCH versus ensemble model across clinical tasks

| | **CONCH Accuracy (%)** | **CONCH Errors (n)** | **Ensemble Accuracy (%)** | **Ensemble Errors (n)** |
|---|---|---|---|---|
| NSCLC Subtyping | 95% | 11 | 95% | 11 |
| CPTAC LUAD EGFR | 58% | 44 | 72% | 30 |
| CPTAC LUAD KRAS | 63% | 39 | 62% | 40 |
| CPTAC LUAD STK11 | 53% | 49 | 75% | 27 |
| CPTAC LUAD TP53 | 71% | 31 | 68% | 34 |
| BERN STAD LAUREN | 55% | 135 | 55% | 138 |
| BERN STAD MSI | 84% | 48 | 87% | 38 |
| BERN STAD N-Status | 69% | 93 | 57% | 132 |
| KIEL STAD LAUREN | 60% | 113 | 58% | 118 |
| KIEL STAD EBV | 92% | 26 | 94% | 19 |
| KIEL STAD MSI | 77% | 73 | 85% | 48 |
| KIEL STAD N-Status | 63% | 119 | 59% | 132 |
| KIEL STAD M-Status | 68% | 104 | 72% | 88 |
| CPTAC BRCA ERBB2 | 83% | 21 | 83% | 21 |
| CPTAC BRCA ESR1 | 69% | 37 | 80% | 24 |
| CPTAC BRCA PGR | 64% | 44 | 67% | 39 |
| CPTAC BRCA PIK3CA | 69% | 37 | 68% | 38 |
| IEO BRCA N-Status | 55% | 203 | 56% | 198 |
| CPTAC CRC Sidedness | 55% | 49 | 53% | 51 |
| CPTAC CRC MSI | 76% | 25 | 67% | 35 |
| CPTAC CRC BRAF | 47% | 56 | 39% | 65 |
| CPTAC CRC KRAS | 65% | 37 | 70% | 32 |
| CPTAC CRC PIK3CA | 47% | 56 | 48% | 55 |
| CPTAC CRC N-Status | 59% | 45 | 60% | 44 |
| DACHS CRC Sidedness | 58% | 1028 | 54% | 1123 |
| DACHS CRC MSI | 89% | 228 | 90% | 199 |
| DACHS CRC BRAF | 66% | 702 | 85% | 314 |
| DACHS CRC KRAS | 41% | 1233 | 36% | 1321 |
| DACHS CRC CIMP | 77% | 515 | 84% | 365 |
| DACHS CRC N-Status | 60% | 941 | 60% | 943 |
| DACHS CRC M-Status | 81% | 334 | 80% | 355 |
| **Sum** | **67%** | **6478** | **69%** | **6076** |

Accuracy and misclassification counts for CONCH and an ensemble across clinical tasks. The Ensemble method combines the average predictions of CONCH, Virchow2, Prov-Gigapath, and DinoSSLPath. The reported numbers are averaged across five cross-validation folds.

## Table S3: Models' architecture overview

| Name | Released | SSL | Architecture | Pretraining Tile size (px) | Patch token size (px) | Magnification | Embed dim | Dataset | Special attributes |
|---|---|---|---|---|---|---|---|---|---|
| CTransPath | Dec 2021 | SRCL | CNN + Swin-Transformer | 1024 | 4 | 20x | 768 | TCGA, PAIP | Mean of all tokens as embbedding |
| DinoSSLPath | Dec 2022 | DINO v1 | ViT-Small | 512 | 16 | 20x, 40x | 384 | TCGA, TULIP | Another DinoSSLPath model with patch token size 8 |
| BiomedCLIP | Mar 2023 | CLIP | ViT-Base | 224 | 16 | diverse | 512 | OpenPath | Vision language model |
| Phikon | Jul 2023 | iBOT | ViT-Base | 224 | 16 | 20x | 768 | TCGA | |
| CONCH | Jul 2023 | iBOT + CoCa | ViT-Base | 256 | 16 | diverse | 512 | MGH, PMC-Path, EDU | Vision language model |
| PLIP | Aug 2023 | CLIP | ViT-Base | 224 | 32 | diverse | 512 | PMC-15M | Vision language model |
| UNI | Aug 2023 | DINO v2 | ViT-Large | 256 & 512 | 16 | 20x | 1024 | BWH, MGH, GTEx | |
| Virchow | Sep 2023 | DINO v2 | ViT-Huge | 224 | 14 | 20x | 1280/2560 | MSKCC | Mean patch tokens added to the tile embeddings |
| Kaiko | Mar 2024 | DINO v2 | ViT-Large | 256 | 14 | 5x,10x,20x,40x | 1024 | TCGA | |
| Prov-GigaPath | May 2024 | DINO v2 | ViT-Giant | 256 | 14 | 20x | 1536 | Providence | LongNet slide encoder to further preprocess tile embeddings |
| Hibou-B | Jun 2024 | DINO v2 | ViT-Base | ? | 16 | ? | 768 | proprietary | |
| Hibou-L | Jun 2024 | DINO v2 | ViT-Large | ? | 16 | ? | 1024 | proprietary | |
| H-optimus-0 | Jul 2024 | DINO v2/iBOT | ViT-Giant | 224 | 14 | 20x | 1536 | proprietary | |
| Virchow 2 | Aug 2024 | DINO v2 (+ ECT | ViT-Huge | 224 | 14 | 5x,10x,20x,40x | 1280/2560 | MSKCC and diverse | Also Virchow2G model with a ViT-Giant architecture |
| | | and KDE) | | | | | | international institutions | |
| Panakeia | - | ? | ViT-Small | 224 | 16 | ? | 384 | proprietary | Specific cancer models only for BRCA and CRC |

## Table S4: Models' pretraining dataset composition

| Name | WSIs (K) | Tiles (M) | Patients (K) | Cancer subtypes | Anatomic sites/Organs | malignant WSIs |
|---|---|---|---|---|---|---|
| CTransPath | 32 | 16 | ~13 | 32 | 25 | 100% |
| DinoSSLPath | 37 | 33 | ? | ? | ? | ? |
| BiomedCLIP | 15,000* | - | ? | ? | ? | ? |
| Phikon | 6 | 43 | 5.6 | 16 | 13 | 100% |
| PLIP | 208* | - | ? | ? | ? | ? |
| CONCH | 1,200* | - | ? | 350 | ? | ? |
| UNI | 100 | 100 | ? | ? | 20 | ? |
| Virchow | 1,488 | 2,000 | 120 | ? | 17 | 38% |
| Kaiko | 29 | 256 | 11 | 32 | 25 | 100% |
| Prov-GigaPath | 171 | 1,385 | 30 | ? | 31 | ? |
| Hibou-B | 1,139 | 512 | 306 | ? | ? | ? |
| Hibou-L | 1,139 | 1,200 | 306 | ? | ? | ? |
| H-optimus-0 | 500 | "several hundreds of millions" | ~333 | ? | ? | ? |
| Virchow2 | 3,135 | 1,700 | 225 | ? | ~175 | 40% |
| Panakeia-BRCA | 5 | 13 | ? | ? | 1 | ? |
| Panakeia-CRC | 1 | 4.5 | ? | ? | 1 | ? |

*image-caption pairs

## Table S5: Proportion of analyzed tissue types in the pretraining data

| Foundation Model | Number represents | Lung | Breast | Stomach | Colon |
|---|---|---|---|---|---|
| Prov-GigaPath | Tissue slides | 45% | 2.7% | 0.7% | 30% |
| CONCH | Image-text pairs | 103k (9.5%) | 65k (5.6%) | 121k (10.4%) | |
| UNI | Tissue slides | 9846 (9.8%) | 3364 (3.3%) | 6705 (6.7%) | 8303 (8.3%) |
| Hibou | Tissue slides in total 112.5k estimated | 2.5k (2.2%) | 12k (11%) | 35k (31%) | |
| Virchow | Tissue slides | 6.1% | 25% | 3.5% | 3.2% |
| Phikon, Kaiko (TCGA) | Patients | 1089 (9.7%) | 979 (8.8%) | 443 (4.0%) | 633 (5.7%) |
| CTransPath | Patients | 1089 (8.4%) | 979 (7.5%) | 443 (3.4%) | 1533 (11.7%) |
| Virchow2 | Tissue slides | ~ 4% | ~ 8% | ~ 2% | ~ 7% |
| Panakeia | Patients | 0 | 4500 (82%) | 0 | 1000 (18%) |

No information available for H-optimus-0, PLIP, BiomedCLIP and DinoSSLPath.

## Table S6: Clinically relevant tasks excluded due to few cases

| STAD | | | | |
|---|---|---|---|---|
| **Marker** | **Value** | **Dataset** | **Cohort** | **Count** |
| NTRK1 | WT | train | TCGA | 321 |
| NTRK1 | MUT | train | TCGA | 5 |
| EBV | negative | test | Bern | 299 |
| EBV | positive | test | Bern | 8 |
| M_STATUS | M0 | test | Bern | 306 |
| M_STATUS | M+ | test | Bern | 1 |
| | | | | |
| **LUAD** | | | | |
| **Marker** | **Value** | **Dataset** | **Cohort** | **Count** |
| BRAF* | WT | train | TCGA | 432 |
| BRAF* | MUT | train | TCGA | 29 |
| BRAF | WT | test | CPTAC | 103 |
| BRAF | MUT | test | CPTAC | 3 |
| MET* | WT | train | TCGA | 441 |
| MET* | MUT | train | TCGA | 20 |
| MET | WT | test | CPTAC | 106 |
| MET | MUT | test | CPTAC | 0 |
| | | | | |
| **CRC** | | | | |
| **Marker** | **Value** | **Dataset** | **Cohort** | **Count** |
| NRAS* | WT | train | TCGA | 529 |
| NRAS* | MUT | train | TCGA | 29 |
| NRAS | WT | test | CPTAC | 100 |
| NRAS | MUT | test | CPTAC | 6 |

*no external validation cohort with at least 10 samples

# Table S7: Patient numbers for individual experiments

| CRC | | | | |
|---|---|---|---|---|
| **Marker** | **Value** | **Dataset** | **Cohort** | **Count** |
| CRC Sidedness | left | train | TCGA | 230 |
| CRC Sidedness | right | train | TCGA | 168 |
| MSI | nonMSIH | train | TCGA | 368 |
| MSI | MSIH | train | TCGA | 61 |
| BRAF | WT | train | TCGA | 450 |
| BRAF | MUT | train | TCGA | 51 |
| KRAS | WT | train | TCGA | 296 |
| KRAS | MUT | train | TCGA | 205 |
| CIMP | nonCIMPH | train | TCGA | 375 |
| CIMP | CIMPH | train | TCGA | 54 |
| PIK3CA | WT | train | TCGA | 377 |
| PIK3CA | MUT | train | TCGA | 124 |
| N_STATUS | N0 | train | TCGA | 318 |
| N_STATUS | N+ | train | TCGA | 238 |
| M_STATUS | M0 | train | TCGA | 417 |
| M_STATUS | M+ | train | TCGA | 76 |
| CRC Sidedness | left | test | Dachs | 1607 |
| CRC Sidedness | right | test | Dachs | 819 |
| MSI | nonMSIH | test | Dachs | 1836 |
| MSI | MSIH | test | Dachs | 210 |
| BRAF | WT | test | Dachs | 1930 |
| BRAF | MUT | test | Dachs | 151 |
| KRAS | WT | test | Dachs | 1397 |
| KRAS | MUT | test | Dachs | 677 |
| CIMP | nonCIMPH | test | Dachs | 1878 |
| CIMP | CIMPH | test | Dachs | 362 |
| N_STATUS | N0 | test | Dachs | 1295 |
| N_STATUS | N+ | test | Dachs | 1085 |
| M_STATUS | M0 | test | Dachs | 1459 |
| M_STATUS | M+ | test | Dachs | 337 |
| CRC Sidedness | right | test | CPTAC | 57 |
| CRC Sidedness | left | test | CPTAC | 51 |
| MSI | nonMSIH | test | CPTAC | 81 |
| MSI | MSIH | test | CPTAC | 24 |
| BRAF | WT | test | CPTAC | 91 |
| BRAF | MUT | test | CPTAC | 15 |
| KRAS | WT | test | CPTAC | 71 |
| KRAS | MUT | test | CPTAC | 35 |
| PIK3CA | WT | test | CPTAC | 87 |
| PIK3CA | MUT | test | CPTAC | 19 |
| N_STATUS | N0 | test | CPTAC | 56 |
| N_STATUS | N+ | test | CPTAC | 54 |
| **STAD** | | | | |
| **Marker** | **Value** | **Dataset** | **Cohort** | **Count** |
| LAUREN | intestinal | train | TCGA | 148 |
| LAUREN | diffuse | train | TCGA | 61 |
| LAUREN | mixed | train | TCGA | 10 |
| EBV | negative | train | TCGA | 300 |
| EBV | positive | train | TCGA | 26 |
| MSI | nonMSIH | train | TCGA | 270 |
| MSI | MSIH | train | TCGA | 56 |
| N_STATUS | N+ | train | TCGA | 225 |
| N_STATUS | N0 | train | TCGA | 97 |
| M_STATUS | M0 | train | TCGA | 289 |
| M_STATUS | M+ | train | TCGA | 21 |
| LAUREN | intestinal | test | Bern | 172 |
| LAUREN | diffuse | test | Bern | 78 |
| LAUREN | mixed | test | Bern | 54 |
| MSI | nonMSIH | test | Bern | 261 |
| MSI | MSIH | test | Bern | 43 |
| N_STATUS | N+ | test | Bern | 205 |
| N_STATUS | N0 | test | Bern | 99 |
| LAUREN | intestinal | test | Kiel | 187 |
| LAUREN | diffuse | test | Kiel | 75 |
| LAUREN | mixed | test | Kiel | 20 |
| EBV | negative | test | Kiel | 302 |
| EBV | positive | test | Kiel | 18 |
| MSI | nonMSIH | test | Kiel | 293 |
| MSI | MSIH | test | Kiel | 27 |
| N_STATUS | N+ | test | Kiel | 222 |
| N_STATUS | N0 | test | Kiel | 98 |
| M_STATUS | M0 | test | Kiel | 259 |
| M_STATUS | M+ | test | Kiel | 61 |
| | | | | |
| **LUAD** | | | | |
| **Marker** | **Value** | **Dataset** | **Cohort** | **Count** |
| EGFR | WT | train | TCGA | 411 |
| EGFR | MUT | train | TCGA | 50 |
| KRAS | WT | train | TCGA | 317 |
| KRAS | MUT | train | TCGA | 144 |
| STK11 | WT | train | TCGA | 394 |
| STK11 | MUT | train | TCGA | 67 |
| TP53 | MUT | train | TCGA | 239 |
| TP53 | WT | train | TCGA | 222 |
| EGFR | WT | test | CPTAC | 72 |
| EGFR | MUT | test | CPTAC | 34 |
| KRAS | WT | test | CPTAC | 74 |
| KRAS | MUT | test | CPTAC | 32 |
| STK11 | WT | test | CPTAC | 88 |
| STK11 | MUT | test | CPTAC | 18 |
| TP53 | MUT | test | CPTAC | 55 |
| TP53 | WT | test | CPTAC | 51 |
| | | | | |
| **NSCLC** | | | | |
| **Marker** | **Value** | **Dataset** | **Cohort** | **Count** |
| NSCLC Subtyping | AC | train | TCGA | 461 |
| NSCLC Subtyping | SCC | train | TCGA | 462 |
| NSCLC Subtyping | AC | test | CPTAC | 106 |
| NSCLC Subtyping | SCC | test | CPTAC | 108 |
| | | | | |
| **BRCA** | | | | |
| **Marker** | **Value** | **Dataset** | **Cohort** | **Count** |
| ERBB2 | negative | train | TCGA | 916 |
| ERBB2 | positive | train | TCGA | 125 |
| ESR1 | positive | train | TCGA | 770 |
| ESR1 | negative | train | TCGA | 271 |
| PGR | positive | train | TCGA | 704 |
| PGR | negative | train | TCGA | 337 |
| PIK3CA | WT | train | TCGA | 687 |
| PIK3CA | MUT | train | TCGA | 336 |
| N_STATUS | N+ | train | TCGA | 554 |
| N_STATUS | N0 | train | TCGA | 468 |
| ERBB2 | negative | test | CPTAC | 106 |
| ERBB2 | positive | test | CPTAC | 14 |
| ESR1 | positive | test | CPTAC | 79 |
| ESR1 | negative | test | CPTAC | 41 |
| PGR | positive | test | CPTAC | 70 |
| PGR | negative | test | CPTAC | 50 |
| PIK3CA | WT | test | CPTAC | 82 |
| PIK3CA | MUT | test | CPTAC | 38 |
| N_STATUS | N+ | test | IEO | 244 |
| N_STATUS | N0 | test | IEO | 207 |

# Table S8: STAMP hyperparameters

| **Hyperparameter** | **Value** |
|---|---|
| Layers | 2 |
| Attention heads | 8 |
| Head activation | GELU |
| Embedding dimension (input) | 384 to 1536 |
| Embedding dimension (reduced) | 512 |
| MLP dimension | 512 |
| Drop path rate (Dropout) | 0 |
| Weight decay | 0.01 |
| Optimizer | AdamW |
| Learning rate | 0.0001 |
| Learning rate schedule | FastAI fit_one_cycle |
| Float precision | Float32 |
| Batch size (training) | 64 |
| Bag size | 512 |
| Batch size (validation/testing) | 1 |
| Training epochs | 32 |
| Early stopping patience | 16 epochs without improvement in AUROC |
| Random seed | Hard-coded |

# Supplementary methods

## Description of foundation models

CTransPath, introduced by Wang et al. in December 2021, is the pioneering Transformer-based unsupervised feature extractor for histopathological images. It integrates a convolutional neural network with a multi-scale Swin Transformer architecture, trained on 15 million patches from 32 cancer subtypes using semantically-relevant contrastive learning, a framework introduced in the same paper based on MoCo v3 [48]. The model was evaluated across multiple tasks, including patch retrieval, patch classification, whole-slide image (WSI) classification, mitosis detection, and colorectal adenocarcinoma gland segmentation [28].

DinoSSLPath, published by Kang et al. in December 2022, employs the DINOv1 framework for SSL [49,50]. DinoSSLPath was pre-trained on 36,666 WSIs, combining TCGA and an internally collected dataset (TULIP). The model uses pathology-specific augmentation techniques, including stain normalization and multi-magnification pretraining at 20× and 40×. DinoSSLPath has been benchmarked on classification tasks and one nuclei instance segmentation task, showing improvements in both label efficiency and dense prediction tasks compared to ImageNet-pretrained baselines.

BiomedCLIP, developed by Zhang et al. in March 2023, is a biomedical vision-language foundation model pretrained on PMC-15M, a dataset of 15 million image-text pairs from PubMed Central [51]. It employs a domain-adapted CLIP framework with PubMedBERT as the text encoder and a ViT-based image encoder. BiomedCLIP excels in cross-modal retrieval, zero-shot classification, and medical visual question answering, achieving state-of-the-art results on diverse biomedical datasets.

Phikon, published by Filiot et al. in July 2023, employs the iBOT framework (image BERT pre-training with Online Tokenizer) for SSL, using MIM and self-distillation [52,53]. The architecture is a ViT-Base model with 80 million parameters, trained on 6,093 WSIs from 16 cancer types, comprising 43 million patches. Phikon was assessed on tile-level and slide-level tasks for subtype, genomic alteration, and overall survival prediction [16].

CONCH, released by Lu et al. in July 2023, is a vision-language model based on CoCa [54], which pretrains an image-text encoder-decoder model using contrastive and captioning losses. For this analysis, only the image encoder was considered. It was pretrained on 16 million image tiles from 21,442 WSIs covering over 350 cancer subtypes. Unlike CTransPath and Phikon, CONCH was trained on proprietary datasets rather than public ones like TCGA and PAIP. The vision-language model is trained by seeking to align image and text modalities in the model's representation space and by predicting the caption corresponding to an image. For this vision-language pretraining, over 1.1 million image-text pairs were used mainly taken from publicly available research articles. Its performance was evaluated on various subtyping, tissue classification, and grading tasks [55].

PLIP, introduced by Huang et al. in August 2023, is a multimodal vision-language foundation model developed for pathology image analysis [56]. Trained on 208,414 pathology image-text pairs from OpenPath—a dataset curated from medical Twitter and other public sources—PLIP employs contrastive learning to align image and text embeddings. The model demonstrates state-of-the-art performance on zero-shot and few-shot classification tasks and also supports image and text-based retrieval, making it a versatile tool for pathology research and education.

UNI, introduced by Chen et al. in August 2023, is notable for being the first model trained on over 100,000 slides. It utilizes DINOv2 for pretraining, incorporating MIM and self-distillation [57]. The training dataset, Mass-100K, was collected from Massachusetts General Hospital and Brigham and Women’s Hospital and consists of over 100 million tissue patches from 20 major tissue types. UNI was tested on the challenging 43-class OncoTree cancer type classification and 108-class OncoTree code classification tasks [21].

Virchow, introduced by Vorontsov et al. in September 2023, stands out as the model trained on the largest dataset to date, with 1.5 million slides from Memorial Sloan Kettering Cancer Center. The model employs DINOv2 for pretraining and features a ViT-Huge architecture with 632 million parameters. Unique to Virchow, the final tile embedding is created using both class tokens and mean patch tokens, doubling the embedding dimension to 2560. It was evaluated on tissue classification and biomarker prediction tasks [23].

Kaiko.ai released a series of pretrained foundation model based on ViT and DINO/DINOv2 [50]. Among these, the ViT-L14 model, trained with DINOv2, was tested in this study. Interestingly, in their own tests, the ViT-B8 trained on DINO performed better on some of the test sets despite using the older SSL method and the smaller ViT-Base architecture. This was attributed to the reduced patch size of eight in comparison to 14 of the ViT-Large. Unlike other recently published foundation models, Kaiko.ai's models were trained on a relatively modest dataset of 29,000 WSIs from TCGA. The performance of these models was assessed on tissue classification tasks using five different datasets [58].

Prov-GigaPath, published by Xu et al. in May 2024, employs a two-stage pretraining approach. Initially, the tile encoder, a ViT-Giant model, is pretrained using DINOv2. Subsequently, a slide encoder contextualizes each tile on the WSI using a LongNet model [59]. Prov-GigaPath was trained on 1.3 billion patches from 170,000 WSIs, sourced from 28 cancer centers within the Providence health network. These WSIs represent over 30,000 patients and encompass 31 major tissue types. Prov-GigaPath was evaluated on mutation prediction using 18 pan-cancer biomarkers and on cancer subtyping tasks, utilizing both TCGA and Providence datasets [22].

PRISM, developed by Shaikovski et al. in May 2024, is a multimodal slide-level foundation model trained on 587,000 WSIs and 195,000 paired clinical reports [60]. Built on Virchow tile embeddings, Prism employs a Perceiver network and BioGPT decoder for zero-shot classification, biomarker prediction, and clinical report generation, achieving state-of-the-art results in low-data scenarios.

Hibou, developed by Nechaev et al. and released in June 2024, includes two versions: Hibou-B and Hibou-L. Hibou-B utilizes a ViT-Base architecture with 86 million parameters and was trained on 510 million tiles. Hibou-L employs a ViT-L architecture trained on 1.2 billion tiles. The training data includes 936,441 H&E and 202,464 non-H&E stained slides from 306,400 individual cases including veterinary biopsies and cytology slides. The performance of the Hibou models was assessed using six datasets for patch-level benchmarks, including tasks such as tissue classification, detection of tumor-infiltrating lymphocytes, and mutation prediction. Additionally, three datasets were used for slide-level benchmarks, focusing on tissue classification [61].

H-optimus-0 was released on GitHub in July 2024 by Saillard et al. It was trained on a proprietary dataset comprising over 500,000 WSIs, from which hundreds of millions of tiles were extracted. Notably, H-optimus-0 features the highest number of WSIs per patient, with an average of 1.5 slides per patient, compared to other models in this study. For instance, Hibou has 3.7 slides per patient, Prov-GigaPath has 5.7 slides per patient, and Virchow has 12.4 slides per patient. H-optimus-0 employs a ViT-Giant architecture with a patch size of 14 and four registers [62]. The model was evaluated on tile-level tissue classification tasks and slide-level biomarker or metastasis prediction tasks [63].

Virchow2, introduced by Zimmermann et al. in August 2024, expands on the Virchow foundation model by scaling its dataset to 3.1 million WSIs from 225,401 patients, sourced from globally diverse institutions, with mixed magnifications and diverse stains [64]. The model employs a ViT-H/14 architecture with 632 million parameters, trained using domain-specific modifications to the DINOv2 framework, including extended-context translation and KDE regularization. Virchow2 demonstrated state-of-the-art performance on 12 tile-level tasks, significantly improving weighted F1 scores on in-domain and out-of-domain benchmarks compared to its predecessor and other foundation models.

MADELEINE, presented by Jaume et al. in August 2024, leverages multistain pretraining to create slide-level embeddings using a Vision Transformer with multihead attention [65]. Trained on 4,211 breast cancer and 12,070 kidney WSIs with various stains, Madeleine demonstrated robust performance across 21 tasks, including morphological subtyping, molecular prediction, and survival analysis.

CHIEF, introduced by Wang et al. in September 2024, is a general-purpose pathology model trained on 60,530 WSIs across 19 anatomical sites using self-supervised tile-level pretraining and weakly supervised slide-level pretraining [66]. CHIEF demonstrated superior generalizability for cancer detection, tumor origin identification, and biomarker prediction.

The proprietary Panakeia models include two cancer-specific ViT-S for CRC and BRCA. The CRC model was trained on 1,000 slides, generating 4.5 million patches, while the Breast model used 5,000 slides, generating 13 million patches.

The collection of histopathology foundation models analyzed in this study is not exhaustive. Additional models published in the past year include BEPH [67], PLUTO [68], RudolfV [69], PathoDuet [70], and models by Campanella et al. [71]. However, these models are either not publicly accessible (PLUTO, RudolfV, Campanella et al.) or have been trained exclusively on TCGA data using the ViT-Base architecture, which renders them less competitive compared to the more recent models evaluated in this study. Furthermore, the publications associated with these models do not offer comparisons with the latest foundation models.

## Comparison of foundation models

In the case of Prov-GigaPath, Xu et al. introduced a slide encoder aimed at analyzing global patterns in WSIs. Previous benchmarking efforts and comparisons by the authors of other foundation models did not evaluate Prov-GigaPath using both tile and slide encoders [61,63,72]. Consequently, we deemed it beneficial to include both versions in this benchmarking study. The results indicate that incorporating the slide encoder does not enhance the performance of the Prov-GigaPath model within a pipeline like STAMP [19]. This is likely because the aggregator model is capable of comprehending slide-level patterns as effectively as the Long-Net model. Thus, it appears feasible to use only the tile encoder in a setup like ours. Similar results were observed for other slide encoders compared to their tile-level counterparts. Using encoded tile embeddings is not beneficial compared to the original tile embeddings; if anything, performance deteriorates. This is likely due to the loss of information, as the encoded embeddings are smaller than the original tile embeddings (**Figure 2G, add. Figure 1A**).

For Virchow, Vorontsov et al. recommended utilizing both the class token (CLS) and the mean patch token (MPT) to create the final tile embedding [23]. This approach doubles the memory requirements for the feature vectors but does not improve performance in our setup. Therefore, it is sufficient to use only the class tokens for Virchow, as is standard practice with the other foundation models in this study. For Virchow2, the same option of using both class tokens and mean patch tokens is available. However, in both cases, we show only the class tokens in the main results, with all versions included for completeness in **Figure S1** and **S10**, and a direct comparison of CLS vs. CLS+MPT for Virchow/Virchow2 in **Add. Fig. 1B**.

CONCH demonstrates exceptional performance despite its relatively modest size as a ViT-Base model. Notably, the vision encoder underwent pretraining on 21,000 WSIs, followed by training the foundation model on over 1.1 million image-text pairs, thus leveraging extensive high-quality training data. Given that the WSIs used for both CONCH and UNI originate from the same source, it is unlikely that there is a qualitative difference between them. The fact that CONCH outperforms UNI, despite being trained on fewer WSIs and utilizing a smaller architecture, underscores the effectiveness of the vision-language approach. These findings suggest that future advancements in histopathological feature extraction may benefit more from the strategic combination of modalities and the utilization of high-quality data rather than merely scaling up model size and data quantity.

## Model Ensembles

Combining the prediction scores of different models and concatenating their feature vectors yielded modest performance improvements compared to the best individual models. Heatmap analyses revealed that different models focus on distinct tissue regions and interpret WSIs differently, suggesting potential benefits in leveraging the strengths of multiple foundation models. The approaches explored in this study, however, were relatively rudimentary. Concatenating feature vectors likely introduced redundancy and created excessively long feature vectors, thereby increasing the risk of overfitting, as per Bellman's curse of dimensionality [73]. This might explain why combining the four best feature vectors resulted in inferior performance compared to CONCH alone. Therefore, it would be interesting to find ways of merging the models without increasing the feature space. To enhance model combination strategies without expanding the feature space, future research could explore dimensionality reduction techniques or the integration of foundation models into a unified framework using merging strategies [74].

## Ablation studies

The challenge of overfitting in machine learning models can often be mitigated by increasing the size of the training cohort. Exclusively for this experiment, we leveraged the DACHS dataset for downstream training, which enabled us to utilize up to 1700 patients across six different tasks. We conducted a 5-fold cross-validation on DACHS and evaluated the models on 11 tasks derived from the TCGA and CPTAC datasets. It is worth noting that models such as CTransPath, Phikon, and Kaiko might possess an inherent advantage in this experimental setup due to their pretraining on TCGA data. Our experiments involved varying the training cohort sizes (100, 200, 400, 850, and 1700 patients) to investigate whether larger embedding vectors yield improved performance with larger training cohorts. We correlated the embedding dimension of each foundation model with the mean AUROC across all tasks and five folds. Contrary to our initial hypothesis, the size of the model's embedding vectors did not consistently influence performance in a straightforward manner. Specifically, smaller virchow-class vectors (1280 dimensions) underperformed compared to virchow vectors (2560 dimensions) with a smaller number of patients. However, this performance disparity diminished when the models were trained on cohorts of 850 or 1700 patients. Similarly, Prov-GigaPath-Slide vectors (768 dimensions) performed worse with smaller patient counts compared to Prov-GigaPath vectors (1536 dimensions), but their performance converged when the full patient cohort was used. A noteworthy observation is that the CONCH model exhibited superior performance with limited training data compared to other models. However, this advantage dissipated as the size of the training cohort increased (**Add. Figure 2**).

## Data diversity in foundation models

An obvious difference between the foundation models lies in the composition of their training data (Table S6). For instance, Virchow's training data consisted of 25% breast tissue, 18.4% skin, and only 6.1% lung. In contrast, UNI's training data predominantly comprised heart and lung tissues, with less than half as many skin and breast cases. Prov-GigaPath's dataset was 45% lung tissue slides, 30% bowel tissue, and only 2.76% breast tissue. Lastly, CONCH placed greater emphasis on GI and lung tissues, with approximately half as much weight given to breast tissue. Given this variability, Virchow's underperformance in lung cases compared to UNI and Prov-GigaPath may reflect the relatively small proportion of lung cases in its training data. However, the fact that CTransPath outperformed Phikon only in BRCA tasks, despite both being trained on the same BRCA cases from TCGA (as PAIP lacks breast tissue), is contrary to this logic. Overall, there is a moderate correlation ($r = 0.41$) between the number of WSIs of a specific tissue type in the pretraining data and relative performance in downstream tasks involving the same tissue type compared to the average performance of all models in the same tasks (Fig S5). While this correlation is not strong, it underscores the importance of considering tissue type diversity when benchmarking foundation models for histopathology.

Due to relying on the STAMP protocol, it was not feasible to include regression or tissue segmentation tasks, which were often part of the original studies for the foundation models. Additionally, expanding the analysis to include more cancer types would be beneficial, as there are noticeable differences in model performance across various cancers. However, given that CONCH consistently performs best across all analyzed cancer types, it is likely a strong candidate for other tasks as well.

# Figures for supplementary methods

## Add. Fig. 1: Comparison of slide encoder models and alternative versions of Virchow and Virchow2

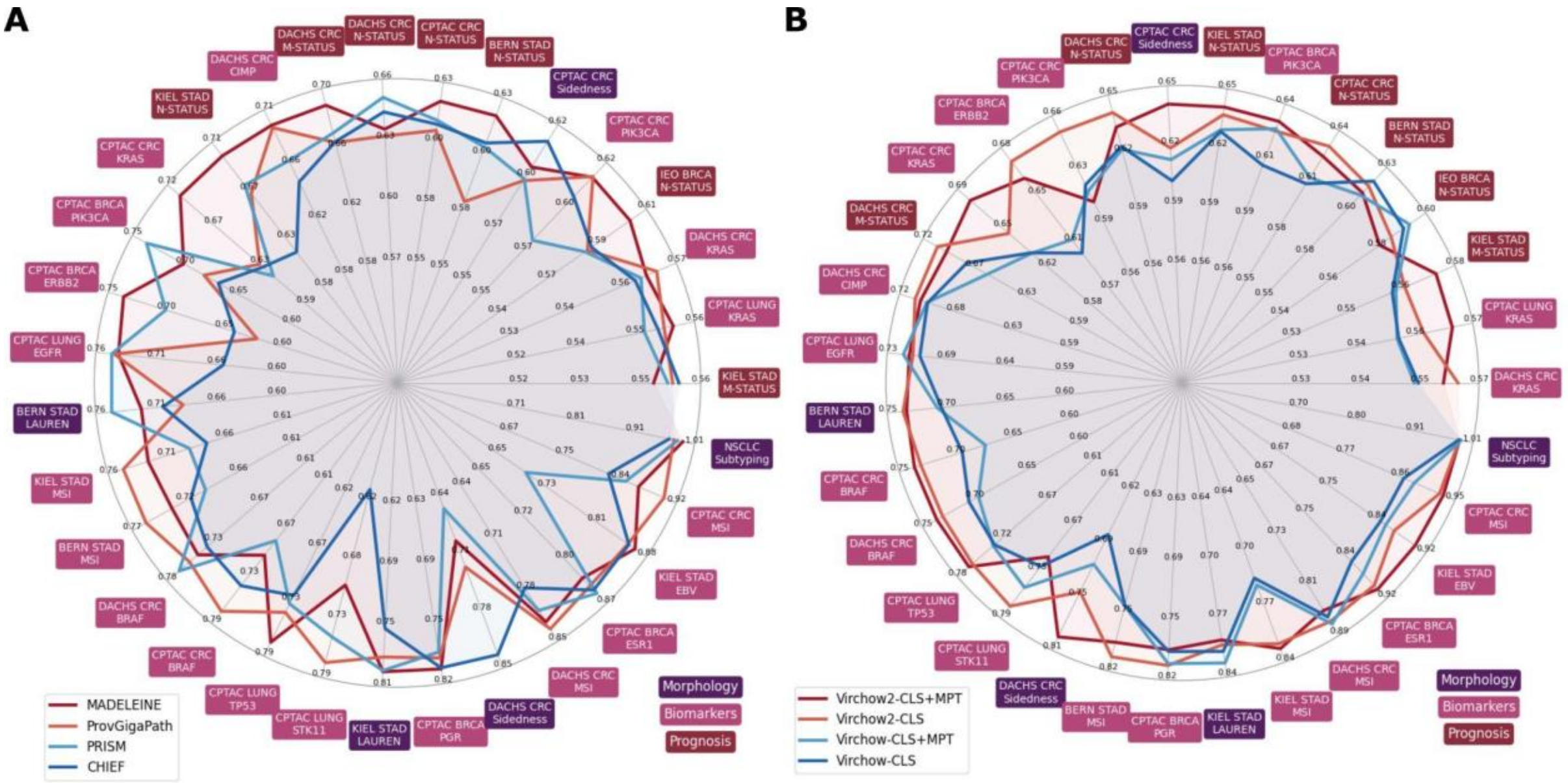

**A**, Average AUROC scores of slide encoders MADELEINE, Prov-GigaPath, PRISM, and CHIEF across all 31 tasks in the benchmark. **B**, Comparison of two versions of Virchow and Virchow2 (class tokens vs. class tokens + mean patch tokens).

# Add. Fig. 2: Experiments with increased downstream training dataset sizes using DACHS as training cohort

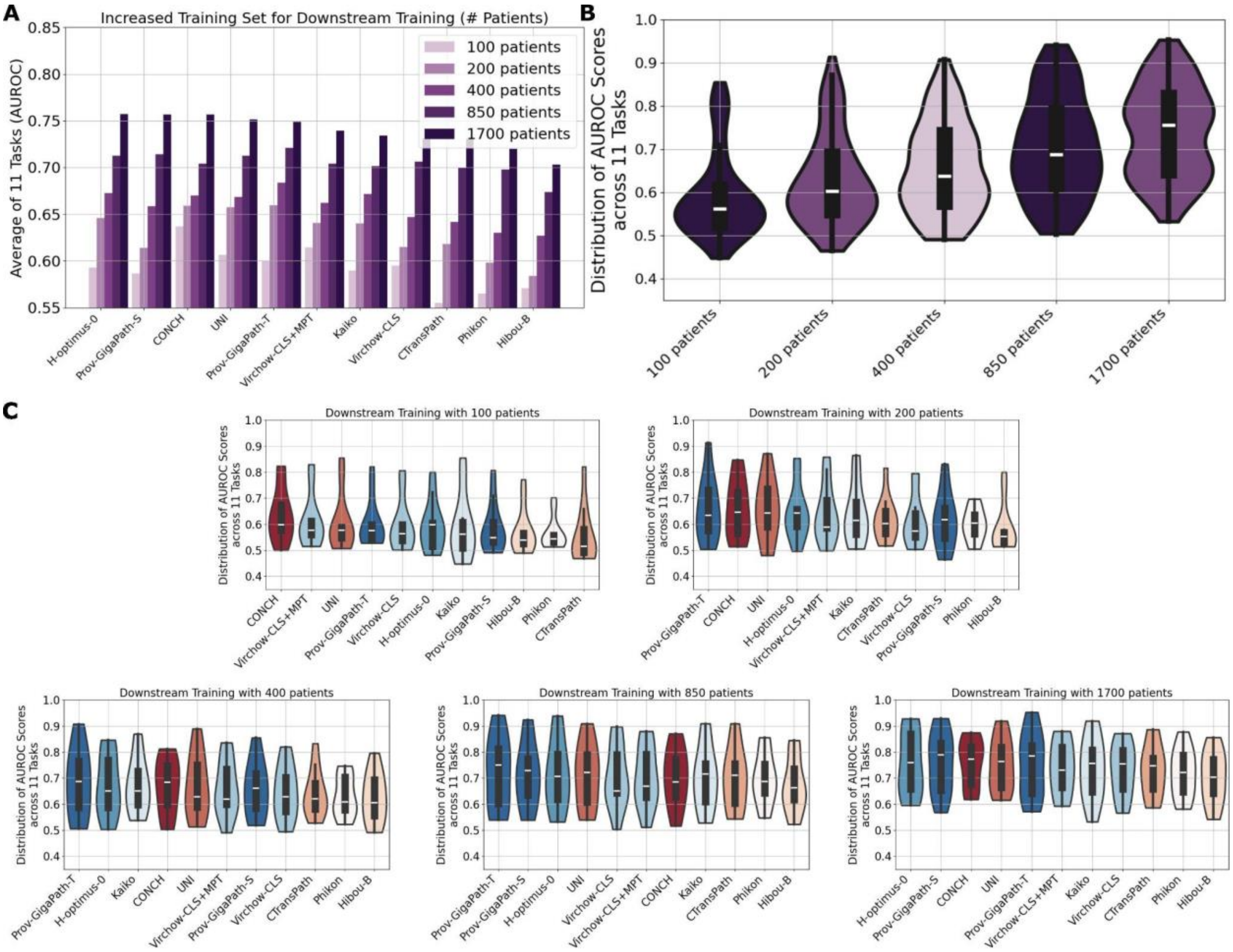

**A**, Average AUROC across five folds on 11 tasks for models trained on a downstream training dataset with 100, 200, 400, 850, or 1700 patients. **B**, Distribution of AUROC scores from all foundation models grouped by downstream training dataset size. **C**, Distribution of AUROC scores for each foundation model individually. For Prov-GigaPath-T, the tile embeddings were used, for Prov-GigaPath-S, the slide encoder was also included. Virchow-CLS only contained the class tokens, Virchow is the version recommended by the authors. Patients were randomly selected from the DACHS cohort, ensuring ground truth was defined for all analyzed tasks. The task "M-Status," was excluded due to insufficient patient numbers. The models were deployed using the CPTAC-CRC cohort and, unlike other experiments, also included the TCGA-CRC cohort. Consequently, Kaiko, CTransPath, and Phikon models might have an advantage as they had prior exposure to TCGA data during pretraining.